\documentclass{nature}

\usepackage{lineno}
\usepackage{amsmath}
\usepackage{amssymb}
\usepackage{graphicx}
\usepackage{xcolor}
\usepackage{mathabx}
\usepackage{hyperref}
\usepackage{comment}
\usepackage{rotating}

\usepackage{xspace}

\newcommand{\Fg}[1]{Figure~\ref{fig:#1}}

\newcommand{\Fgs}[2]{Figures\ \ref{fig:#1} and \ref{fig:#2}}

\newcommand{\eq}[1]{Eq.~(\ref{eq:#1})\xspace}
\newcommand{\Eq}[1]{Equation~(\ref{eq:#1})\xspace}
\newcommand{\eqs}[2]{Eqs.\ (\ref{eq:#1}) and (\ref{eq:#2})}

\newcommand{\tb}[1]{Table~\ref{tab:#1}\xspace}

\newcommand{\taud}{\tau_{\rm d}}
\newcommand{\vr}{v_{\rm r}}
\newcommand{\auyr}{\rm \ AU \ Myr^{-1 }}
\newcommand{\Sigmap}{\Sigma_{\rm g}}
\newcommand{\Msunyr}{ \rm \ M_{\odot} \ yr^{-1}}

\newcommand{\Msyr}{ \  \rm M_\odot \, yr^{-1} }

\newcommand{\Me}{ \ \rm M_\oplus}

\newcommand{\Omegak}{\Omega_{\rm K}}
\newcommand{\Omegap}{\Omega_{\rm p}}

\newcommand{\qp}{q_{\rm p}}
\newcommand{\qd}{q_{\rm d}}

\newcommand{\alphat}{\alpha_{\rm t}}

\makeatletter
\let\saved@includegraphics\includegraphics
\AtBeginDocument{\let\includegraphics\saved@includegraphics}
\renewenvironment*{figure}{\@float{figure}}{\end@float}
\makeatother

\title{Early Solar System instability triggered by dispersal of the gaseous disk} 

\author{Beibei Liu$^{1,2 \star}$,  Sean N. Raymond$^{3}$, and Seth A. Jacobson$^{4}$ }

\begin{document}

\maketitle

\begin{affiliations}
 \item Zhejiang Institute of Modern Physics, Department of Physics \& Zhejiang University-Purple Mountain Observatory Joint Research Center for Astronomy, Zhejiang University, 38 Zheda Road, Hangzhou 310027, China
 \item Department of Astronomy and Theoretical Physics, Lund Observatory,  Box 43, SE--22100, Sweden
 \item Laboratoire dAstrophysique de Bordeaux, Univ. Bordeaux, CNRS, B18N, all{\'e}e Geoffroy Saint-Hilaire, 33615 Pessac, France
 \item Department of Earth and Environmental Sciences, Michigan State University, East Lansing, MI 48824, USA
 \item[*] To whom correspondence should be addressed; Email: {bbliu@zju.edu.cn}
\end{affiliations}

\begin{abstract}
The Solar System's orbital structure is thought to have been sculpted by an episode of dynamical instability among the giant planets\cite{Tsiganis2005, Morbidelli2007b, Batygin2012, Nesvorny2018a}.
However, the instability trigger and timing have not been clearly established\cite{Levison2011,Nesvorny2018b,Mojzsis2019,Quarles2019,deSousa2020}.  
Hydrodynamical modeling has shown that while the Sun's gaseous protoplanetary disk was present the giant planets migrated into a compact orbital configuration in a chain of resonances\cite{Morbidelli2007b, Pierens2014}.
Here we use dynamical simulations to show that the giant planets' instability was likely triggered by the dispersal of the gaseous disk.
As the disk evaporated from the inside-out, its inner edge swept successively across and dynamically perturbed each planet's orbit in turn.
The associated orbital shift caused a dynamical compression of the exterior part of the system,  ultimately triggering instability.
The final orbits of our simulated systems match those of the Solar System for a viable range of astrophysical parameters.
The giant planet instability therefore took place as the gaseous disk dissipated, constrained by astronomical observations to be a few to ten million years after the birth of the Solar System\cite{Williams2011}.
Terrestrial planet formation would not complete until after such an early giant planet instability\cite{Jacobson2014, Kleine2017}; the growing terrestrial planets may even have been sculpted by its perturbations, explaining the small mass of Mars relative to Earth\cite{Clement2018}.
\end{abstract}

We modeled the dynamical consequences of the dispersal of the Sun's gaseous disk.
Stellar photo-evaporation dominates the mass-loss during this advanced phase, causing the disk to dissipate from the inside-out\cite{Alexander2014, Ercolano2017}.  While planets embedded in the disk feel ``two-sided'' gravitational torques from both the interior and exterior parts of the disk, planets at the disk's inner edge only interact with the gas exterior to their orbits.
As a result of these larger ``one-sided'' torques,  a planet below the mass threshold for opening a gap will stop migrating inward at the disk inner edge\cite{Masset2006, Romanova2019}.
If the inner edge itself moves outward due to disk dispersal, then the planet may subsequently migrate outward along with it (Methods).
This mechanism is termed ``rebound'', and was first applied in the context of the magnetospheric cavity on sub-AU scales to explain the architecture of close-in super-Earth planets\cite{Liu2017a, Liu2017b}.

\Fg{illustration} demonstrates an example simulation of a dynamical instability triggered by the disk's dispersal.
The expanding edge of the inner disk cavity does not affect all planets equally.
Because Jupiter is sufficiently massive enough to open a deep gap around its horseshoe region, the corresponding corotation torque diminishes and the rebound is quenched (Methods).
Jupiter then simply enters the cavity as the inner disk edge sweeps by.
The one-sided torque is strong enough to expand Saturn's orbit outward when the disk edge approaches Saturn at $t {=} 0.6$ Myr  (\Fg{illustration}), moving Jupiter and Saturn out of their shared resonance.
As Saturn migrates outward with the expanding cavity, the spacing between the orbits of the outer planets is compressed.
The eccentricities of the ice giants increase due to this dynamical compression.
Saturn is left behind and enters the cavity at $9$ AU when $t{=} 0.65$ Myr.
Meanwhile, the innermost ice giant planet becomes so dynamically excited that its orbit crosses Saturn's, and the two planets undergo a close gravitational encounter.
This triggers a dynamical instability, and the system becomes chaotic: the third ice giant is scattered outward, while the innermost ice giant is eventually ejected into interstellar space at $t{=} 0.85$ Myr after a series of close encounters with Jupiter. The planets' final orbits are close to those of the present-day Solar System giant planets.

Such a rebound-triggered instability is consistent with the Solar System's orbital architecture.
To demonstrate this, we conducted more than 14,000 numerical simulations like the one from \Fg{illustration} varying three different aspects of the initial conditions (Extended Data \tb{IC}).
First, we tested a wide range of plausible starting configurations for the number of ice giants ($2$, $3$, or $4$) and their initial orbital resonant states.
Second, we used a Monte Carlo method to test the effects of important disk parameters --- the onset mass-loss rate $\dot M_{\rm pho}$, the disk dispersal timescale $\taud$, and the expansion rate of the inner cavity $\vr$ --- across the full range of astronomically-relevant values.
Third, we ran each simulation twice: once including the effect of inside-out disk dissipation (i.e. with rebound) and once assuming the disk dissipates smoothly at all radii (i.e. without rebound).  
As a basic check, we used two system-level indicators to test whether our simulated systems are consistent with the global properties of the Solar System: the (normalized) angular momentum deficit AMD, a measure of the dynamical excitation of the system, and the radial mass concentration statistic RMC, a measure of the orbital spacing of the system.

When the rebound effect is included in our simulations, the surviving planetary systems fill the AMD-RMD phase space that matches the Solar System (\Fg{5planet}). That space is mostly empty when rebound is not included because dynamical instabilities are much less frequent.
More than $90\%$ of systems starting in $3$:$2$ resonances went unstable when rebound was included but only $39\%$ when rebound was ignored.
Likewise, $78\%$ of systems with a chain of $2$:$1$ (Jupiter and Saturn) and $3$:$2$ resonances went unstable when rebound was included vs. $31\%$ when it was ignored.
The rebound-triggered instability occurs across all astronomically-relevant disk parameter values (Methods).  In these simulations, we had adopted a moderately viscous disk in which Saturn did not open a deep gap.  
However, the rebound mechanism also generates instability in low-viscosity environments where Saturn is above the gap-opening mass.
In that case, the ice giants' scattering propagates to the gas giants, triggering a system-wide instability at a rate that is only modestly lower than in our fiducial simulations (Methods).

In previous studies\cite{Tsiganis2005,Gomes2005,Levison2011,Nesvorny2012,Quarles2019,deSousa2020}, a primordial planetesimal disk typically contained $20{-}30 \Me$ within $30$ AU and played a central role in triggering the instability.
In our model, the gas disk is the instability trigger, yet interactions with a putative outer planetesimal disk would further spread out the giant planets' orbits and decrease their eccentricities and inclinations.
After the gas disk was fully dissipated, we extended a subset of simulations in a gas-free environment for another $100$ Myr including an outer planetesimal disk containing a total of $5$, $10$, or $20\Me$.
In an example with four giant planets (\Fg{plt_planet}a,b), the rebound-driven instability leaves the system in a configuration that is more compact than the real one.
Yet, during the planetesimal disk phase ($t{>}10$ Myr) the orbital radius of Uranus and Neptune increased, and the eccentricities of all planets were damped, resulting in a configuration closer to that of the Solar System.
An example starting with five giant planets with an outer planetesimal disk of $5 \Me$ followed a similar evolutionary path (\Fg{plt_planet}c,d).
In dynamical terms, the rebound-triggered instability increases a giant planet system's level of orbital excitation (and its AMD) and decreases its degree of radial concentration (and RMC), whereas later interactions with the planetesimal disk tend to decrease both the AMD and RMC.

The final system architectures provide a better match to the Solar System when planetesimal disks were included (\Fg{aei}).
One challenge for our simulations is adequately exciting Jupiter's eccentricity to its present value of $0.046$.
This is a systematic problem in simulations of the instability\cite{Tsiganis2005,Nesvorny2012,Batygin2012,Clement2018,Clement2021}.
A possible solution is that Jupiter's orbit was already modestly eccentric at the tail end of the gaseous disk phase\cite{Pierens2014, Clement2021}.
We do not attempt to explain the Kuiper Belt's architecture in this work, since the triggering mechanism is not the central aspect for establishing these small body populations.
The chaos of the instability erases the dynamical memory of the initial triggering--a defining feature of chaos--and the dissipating gas in the dispersal phase only plays a minor role in damping the random velocities of small bodies once they get excited.
Thus, results regarding existing models of small body evolution after giant planet instability hold regardless of the triggering mechanism.

A rebound-triggered instability at the time of disk dispersal fills an important gap in Solar System chronology.
Observations of the frequency of disks in star clusters of different ages find that the typical disk lifetime is a few to ten Myr\cite{Williams2011}. 
The giant planet instability was initially invoked as a delayed event to coincide with the ``late heavy bombardment''\cite{Gomes2005}. 
However, recent re-appraisal of the cosmochemical constraints indicates that there was likely no late spike (``terminal cataclysm'') in the bombardment rate\cite{Zellner2017}. 
Instead, constraints from a binary Jupiter Trojan\cite{Nesvorny2018b} and ages of meteoritic inclusions\cite{Mojzsis2019} suggest that the instability took place no later than $\sim$20-100 Myr after the birth of the Solar System.
An instability within 10 Myr would have perturbed the final assembly of the terrestrial planets, and an early instability may explain a number of features of the inner Solar System including the large Earth-to-Mars mass ratio and the dynamical excitation of the asteroid belt\cite{Clement2018}. 

Our model provides a generic trigger for dynamical instability linked with the observed timescale for disk dispersal\cite{Williams2011}.
Early models relied on fine-tuning the distance between the ice giants and outer planetesimal disk or the degree of self-interaction between planetesimals to match the assumed late timing of the instability\cite{Gomes2005, Levison2011}.
More recent studies with more self-consistent outer planetesimal disks systematically find shorter instability timescales, but with broad distributions that extend to ${\sim} 100$ Myr and an uncertainty in the triggering mechanism itself\cite{Quarles2019,deSousa2020}.
Given its robustness to disk parameters (Extended Data \Fgs{appendix_diskparameter4planets}{appendix_diskparameter5planets}), an early, rebound-triggered instability is essentially unavoidable for the Solar System.
Such a trigger also implies that the planets' final architecture depends only weakly on the mass of the outer planetesimal disk (\Fg{aei}), which may have been less massive than previously thought.
For instance, the cratering record on outer Solar System moons does not appear to match predictions from models of the collisional evolution of the massive primordial Kuiper Belt needed to trigger the giant planets' instability\cite{Zahnle2003, Singer2019}.

The rebound effect may explain why dynamical instabilities appear to be nearly ubiquitous in exoplanetary systems\cite{Raymond2018}.
The broad eccentricity distribution of giant exoplanets can be matched if 75-95\% of all giant planet systems we see are the survivors of dynamical instabilities\cite{Raymond2010}.
While rebound only affects planets with masses below the gap-opening mass (such as ice giants), microlensing studies find that such planets are extremely common\cite{Suzuki2016}.
The instabilities in ice giant systems spread to nearby gas giants at a high probability (Methods).
Likewise, the orbital period ratio distribution of close-in small planets is consistent with the vast majority of systems having undergone dynamical instabilities\cite{Pu2015}.
Thus, the rebound effect during disk dispersal may be a nearly universal process affecting not just our Solar System but planetary systems across the Galaxy.  


\begin{addendum}

\item[Correspondence and requests for materials] 
Correspondence and requests for materials should be addressed to B.~L. (email: bbliu@zju.edu.cn).

\item[Acknowledgements] B.~L.  is supported by the start-up grant of the Bairen program from Zhejiang University, National Natural Science Foundation of China (No. 12173035 and 12111530175), the Swedish Walter Gyllenberg Foundation and the European Research Council (ERC Consolidator Grant 724687- PLANETESYS). S.~N.~R.  is grateful to the CNRS's PNP program.

\item[Author contributions] 
S.~N.~R. and S.~A.~J. proposed this idea and initiated the collaboration. B.~L. examined the feasibility and conducted numerical simulations. S.~N.~R. drafted the manuscript. All authors contributed to analyzing and discussing the numerical results, editing, and revising the manuscript. 

\item[Author information]  
The authors declare that they have no competing interests. 
\end{addendum}


\clearpage

\begin{figure}
    \includegraphics[width=16cm]{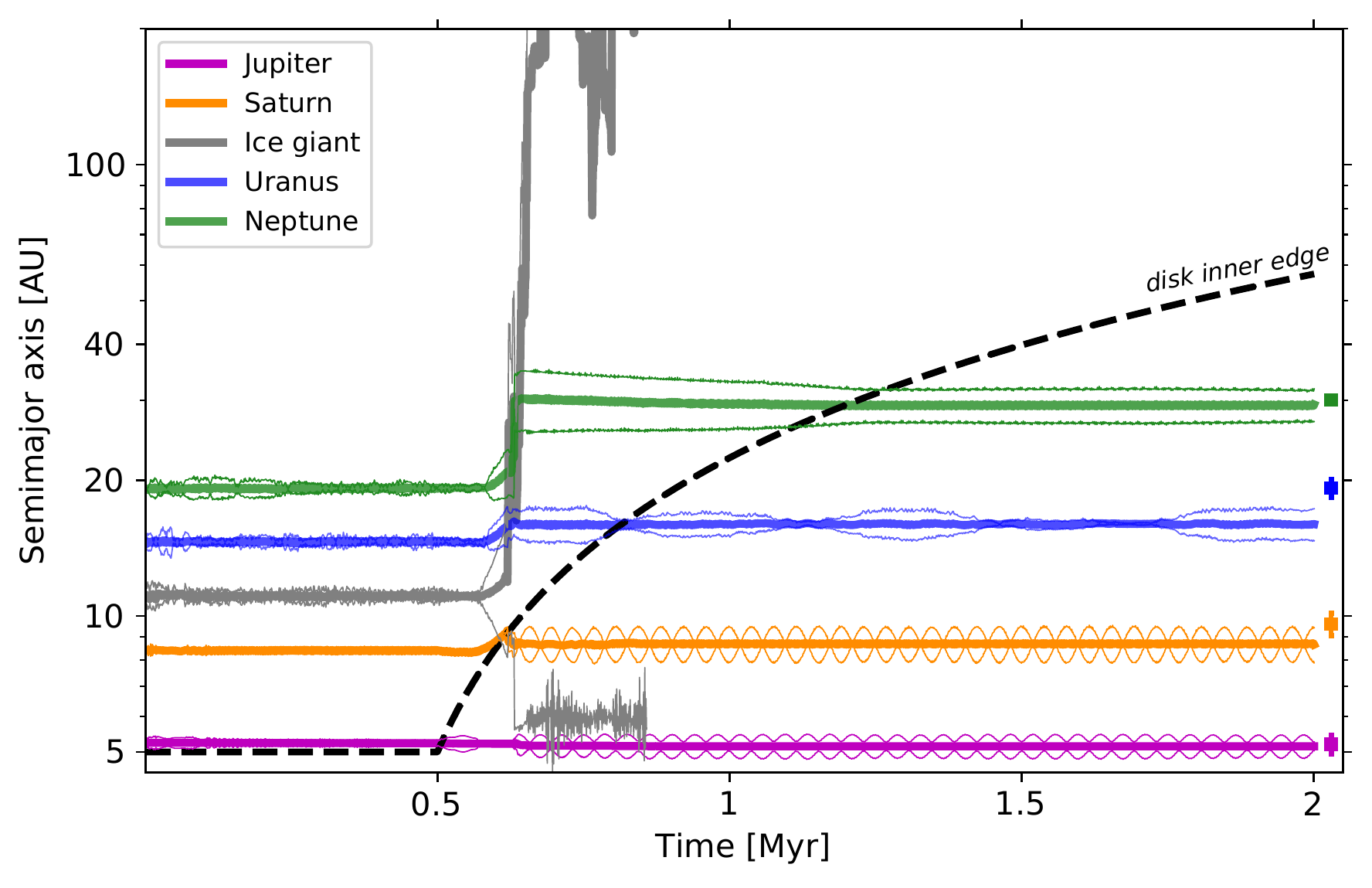} 
    \caption{\textbf{An early dynamical instability triggered by the dispersal of the Sun's protoplanetary disk.}
    The initial system consisted of five giant planets: Jupiter, Saturn, and three $15 \ M_{\oplus}$ ice giants, one of which was ejected into interstellar space during the instability.
    The curves show the orbital evolution of each body including its semimajor axis (thick), perihelion and aphelion (thin).
    The black dashed line tracks the edge of the disk's expanding inner cavity.
    We do not follow the early evolution through the entire gas-rich disk phase, so the onset of disk dispersal is set arbitrarily to be $0.5$ Myr after the start of the simulation.
    The semimajor axes and eccentricities of the present-day giant planets are shown at the right, with vertical lines extending from perihelion to aphelion.
    The adopted disk parameters are: $\dot M_{\rm pho} {=}4 \times 10^{-10} \Msunyr$, $\taud{=}8.6\times 10^{5}$ yr, and $\vr{=} 35 \auyr$.} 
    \label{fig:illustration}
\end{figure}

\begin{figure}
    \includegraphics[width=14cm]{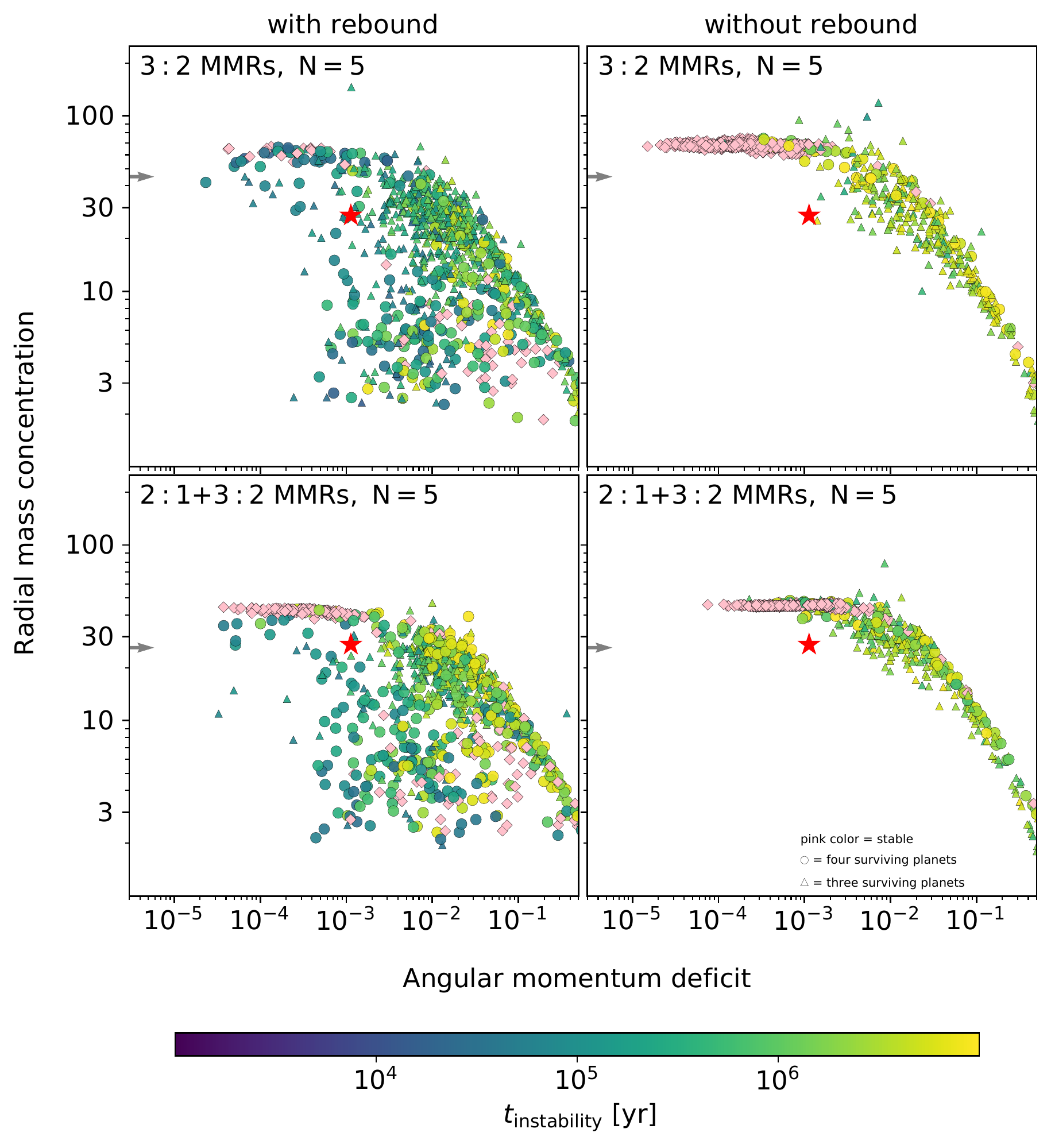}
    \caption{\textbf{Metrics for surviving planetary systems of a subsample of our simulations in matching the Solar System.}
    The simulations on the left included the rebound effect and those on the right did not.
    Each simulation started with our four present-day giant planets plus one additional ice giant planet.
    In the top panels, the giant planets were initially placed in a chain of $3$:$2$ orbital resonances.
    In the bottom panels, Jupiter and Saturn were in a $2$:$1$ resonance and other neighboring planet pairs were in $3$:$2$ resonances.
    Each symbol represents the outcome of a given simulation at $t{=}10$ Myr.
    The color indicates the timing of the instability after the start of gas disk dispersal; pink systems did not undergo an instability (no collision and/or ejection).
    Diamonds, circles, and triangles correspond to systems with five, four, and three or fewer surviving planets, respectively.
    The arrow gives the initial radial mass concentration of the system.
    The Solar System is marked as a red star for comparison.
    Comparable figures presenting different subsamples starting from different orbital configurations are included as Extended Data Figures \ref{fig:appendix_dynamics4planets}, \ref{fig:appendix_dynamics5planets} and \ref{fig:appendix_dynamics6planets}.}
    \label{fig:5planet}
\end{figure}

\begin{figure}
\includegraphics[width=16cm]{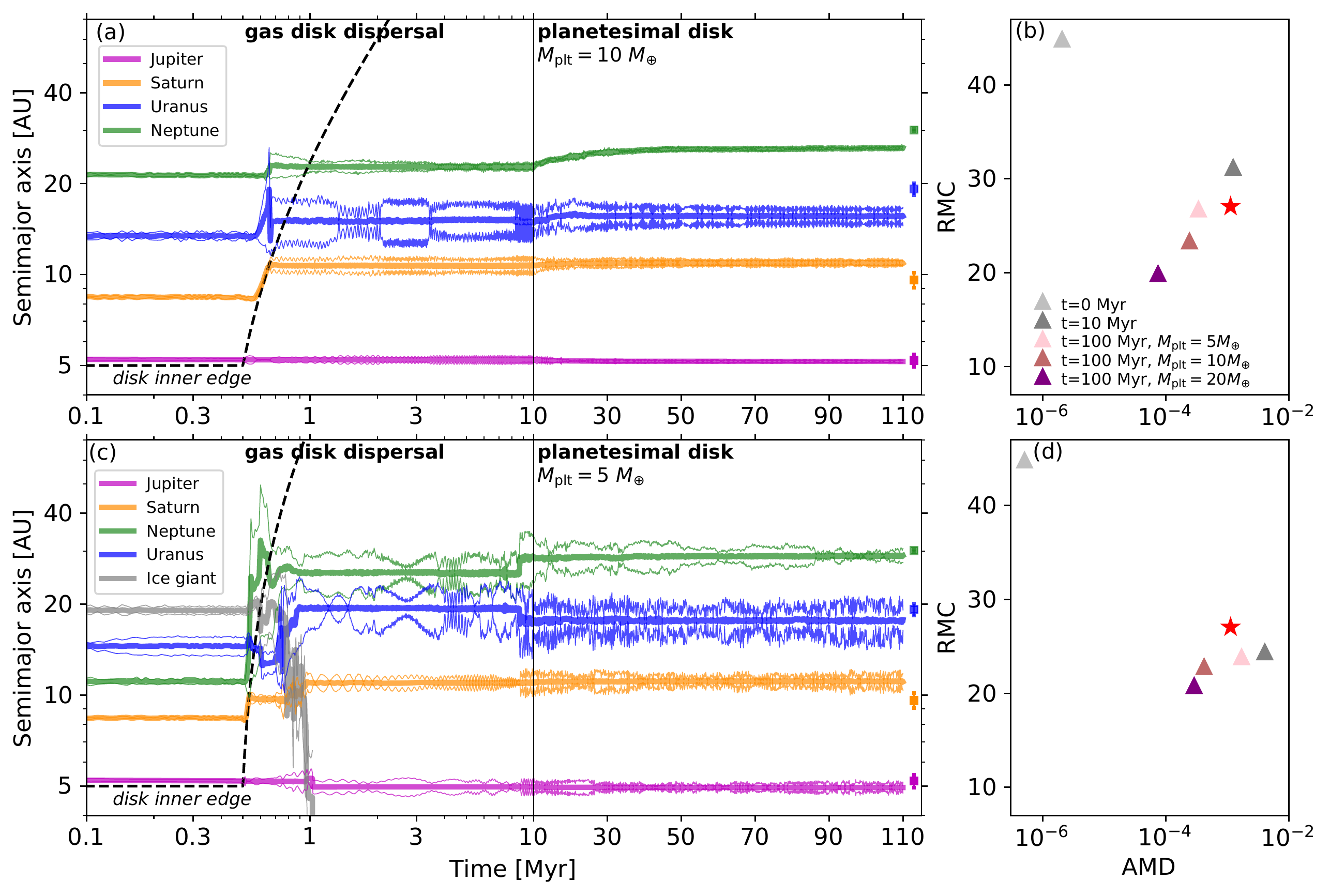}
\caption{\textbf{Dynamical evolution of giant planets in both gas disk dispersal phase and gas-free, planetesimal disk phase. } 
The initial system consisted of four giant planets in $2$:$1$ resonances (upper) or five giant planets in a combined  $2$:$1$ and $3$:$2$ resonances (lower).
The left panels show the orbital evolution of each body including its semimajor axis, perihelion, and aphelion. 
The black dashed line tracks the edge of the disk's expanding inner cavity.  The onset of disk dispersal is set arbitrarily to be $0.5$ Myr after the start of the simulation.
The planetesimal disks of $10\Me$ and $5\Me$ are implemented after $10$ Myr in the above two configurations. 
The right panels provide the corresponding system's radial mass concentration (RMC) and normalized angular momentum deficit (AMD) at $t{=}0$ yr, $10$ Myr, and $100$ Myr with a planetesimal disk of $5 \Me$  (pink),  $10 \Me$ (brown), and $20\Me$ (purple), respectively.
Solar System is marked as a red star for comparison.
}
\label{fig:plt_planet}
\end{figure}

\begin{figure}
\includegraphics[width=16cm]{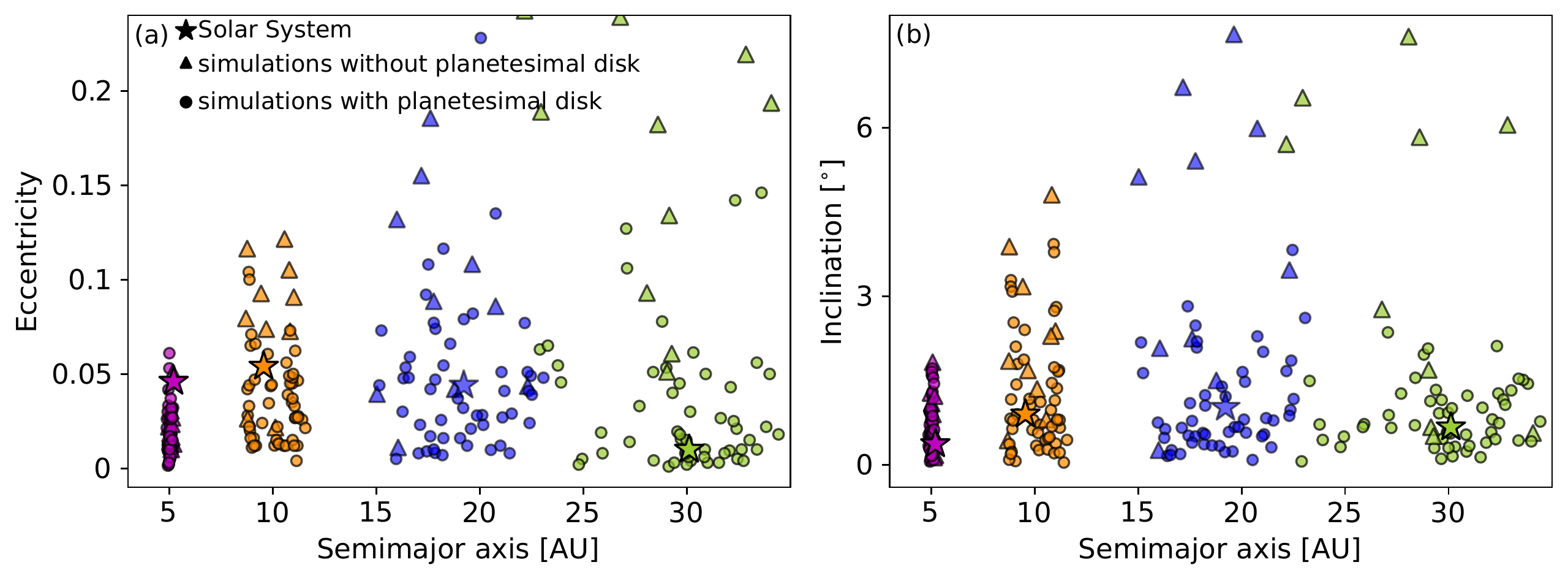}
\caption{\textbf{Final orbits of the giant planets in simulations that included both the gaseous disk and an outer planetesimal disk.} 
Only the simulations that finished with four planets are shown.
Simulations with and without planetesimal disks are plotted in circles and triangles, respectively.
The Solar System is marked as a star.  
}
\label{fig:aei}
\end{figure}

\clearpage
\begin{methods}
\subsection{Disk model.}

The evolution of a protoplanetary disk is driven by internal gas viscous stresses and external stellar UV/X-ray photoevaporation\cite{Alexander2014, Ercolano2017}.
At early times the influence of stellar photoevaporation is negligible and the disk evolves viscously in a quasi-steady state such that $ \dot M  {=}3 \pi \nu \Sigma$, where $ \dot M$ is the gas disk accretion rate, $\Sigma$ is the gas surface density, $\nu{=} \alpha c_{\rm s}H$ is the disk  viscosity\cite{Shakura1973}, $\alpha$ is the viscous efficiency parameter, and $c_{\rm s}$ and $H$ are the gas disk sound speed and scale height, respectively.
At late times, after the disk accretion rate $\dot M$ drops below the stellar photoevaporation mass-loss rate $\dot M_{\rm pho}$, disk dissipation is dominated by the photoevaporative wind originating from stellar high-energy radiation\cite{Alexander2006,Owen2011}.

When the thermal energy of the disk gas is greater than its gravitational bounding energy beyond a threshold disk radius $R_{\rm in}$, the disk gas escapes as a wind.
Gas interior to $R_{\rm in}$ is rapidly accreted on the local viscous timescale. 
Hence, an inner disk cavity is opened and the main outer disk is optically thin to the direct radiation from the central star\cite{Alexander2006}.
As such, the photoevaporating mass-loss rate is  given by\cite{Alexander2014}
\begin{equation}
 \dot M_{\rm pho} \simeq   10^{-9} \left(\frac{  \Phi}{10^{41} \rm \ s^{-1}}  \right)^{1/2}\left(\frac{R_{\rm in}}{ 5 \rm \ AU}  \right)^{1/2} \ \rm M_{\odot}\ yr^{-1},
     \label{eq:mdot_pho}
\end{equation}
where the ionizing flux of the young central star $ \Phi$ is order of $10^{41}{-} 10^{42} \rm \ s^{-1}$   and the initial size of inner cavity $R_{\rm in}$ is $1{-}5$ AU.  An initially smaller size cavity does not qualitatively change our results,  and the determining factor is $\dot M_{\rm pho}$ when the inner disk edge sweeps at Jupiter's location.
The mass-loss rate depends on the primary stellar incident spectrum (EUV, FUV, or X-ray), grain species, abundance, and disk chemistry.
The fiducial onset mass-loss rate $\dot M_{\rm pho}$ of $ 10^{-9} \Msyr$ in \eq{mdot_pho} is taken from  EUV-driven stellar radiation models\cite{Alexander2006}.  
If the photoevaporation is instead dominated by X-ray radiation, $\dot M_{\rm pho}$  can be  higher\cite{Owen2011}.
We treat $\dot M_{\rm pho}$ as a free parameter and vary it from $3 {\times} 10^{-9}$ to $3 {\times }10^{-10} \Msyr$ in the parameter study.  

In this photoevaporation-driven disk dispersal phase, the gas surface density can be written:   
\begin{equation}
\Sigma (t) = \Sigma_{\rm pho} \exp\left[-\frac{t}{\taud}\right] = \frac{\dot M_{\rm pho}}{3\pi \nu} \exp\left[-\frac{t}{\taud}\right],
 \label{eq:Sigma}
\end{equation}
where  $\Sigma_{\rm pho}$ is the gas surface density when the stellar photoevaporation becomes dominant and $\taud $ is the gaseous disk dispersal timescale.
Noticeably, two different timescales are related to gas disk dissipation.
First, the disk lifetime is inferred to be $1{-}10$ Myr with a median value of $3$ Myr\cite{Haisch2001}.
Second, since only a small fraction of young stars have evolved disks caught in the transition between disk-bearing and disk-less states, disk clearing is expected to be very rapid, typically an order of magnitude shorter than the disk lifetime\cite{Luhman2010, Koepferl2013}.
Based on these observational constraints, it is reasonable to assume that $\taud$ is in a range between $0.1$ and $1$ Myr. 

 We assume an optically thin, flaring disk in this dispersing phase\cite{Hayashi1981}, and the corresponding temperature and aspect ratio are:
\begin{equation}
  T  = \left( \frac{L_{\star}}{ 16 \pi \sigma_{\rm SB} r^2} \right)^{1/4} =   T_{\rm in} \left(\frac{r}{ R_{\rm in}} \right)^{-1/2} \quad \text{and} \quad
    h =  \frac{H}{r} 
    = \sqrt{\frac{ k_{\rm B} T}{\mu m_{\rm p}} \frac{r}{G M_{\star}}}  =   h_{\rm in} \left(\frac{r}{ R_{\rm in} }  \right)^{1/4},   
  \label{eq:h}
 \end{equation}
 where $M_{\star}$ and $L_{\star}$ are the stellar mass and luminosity, $r$ is the disk radial distance, $G$ is the gravitational constant, $\sigma_{\rm SB}$ is the Stefan-Boltzmann constant, $k_{\rm B}$ is the Boltzmann constant, $\mu$ is the gas mean molecule weight, and $m_{\rm p}$ is the proton mass.  
 Ref.\cite{Hayashi1981} calculated that $h{=}0.033$ at $1$ AU by adopting $L_{\star} {= }1  \ L_{\odot}$.
 However, during the pre-main-sequence (stellar age $<$ a few Myr), the Sun underwent gravitational contraction and was more luminous (${\sim}1{-}10 \ L_{\odot}$, along the Hayashi track) than its main-sequence stage.
 Hence, we choose a higher luminosity for a solar-mass pre-main-sequence star of $L_{\star} {=} 4 \ L_{\odot}$ and assume it remains constant during the relatively rapid disk dispersal phase.
 In this circumstance, the disk aspect ratio is $0.04$ at $1$ AU.
 We adopt $R_{\rm in}{=}5$ AU and therefore $h_{\rm in}{=}0.06$ and $T_{\rm in}{=} 180\ \rm K$, where the subscript indicates the quantity evaluated at $R_{\rm in}$ when the stellar photoevaporation dominates.
 Based on the quasi-steady state assumption, we rewrite the gas surface density as: 
\begin{equation}
    \Sigma(t) = \frac{\dot M_{\rm pho}}{3 \pi \nu} \exp\left[-\frac{t}{\taud}\right]  = \Sigma_{\rm in}  \left(\frac{r}{ R_{\rm in}}  \right)^{-1} \exp\left[-\frac{t}{\taud}\right], 
     \label{eq:sig}
\end{equation}
where $ \Sigma_{\rm in} {=} 4  \ \rm g \ cm^{-2} $ for the fiducial disk parameters of $ \dot M_{\rm pho} {=} 10^{-9}  \ \rm M_{\odot}\ yr^{-1}$ and $\alpha {=} 0.005$.

The cavity spreads from inside-out as disk gas disperses.
 We assume that the cavity expands with a constant speed $\vr$ for simplicity.
The cavity expansion rate should be principally determined by stellar ionizing flux, size and mass of the disk, and hence, not a fully independent variable.
Nonetheless, when the dispersal time and disk mass are fixed, $\vr$ directly reflects the size of the disk.
To a first order approximation, for a disk with an outer edge $R_{\rm out}$ of $30$ AU and dispersal timescale of $0.5$ Myr, $\vr {\sim}R_{\rm out}/\taud {= }60 \auyr$.
Due to the uncertainties in $R_{\rm out}$ and $\taud$, $\vr$ could plausibly span two orders of magnitude from $10$ to $10^{3} \auyr$. 
We neglect the effect of disk self-gravity in this study, which might induce secular resonances with the gas giant planets to sweep through the inner Solar System and cause the orbital excitation of asteroids and terrestrial planets--understanding how the rebound instability impacts the inner Solar System is a focus for future work.

In brief, the final photoevaporation-driven disk dispersal can be described by three key model parameters: the onset mass-loss rate when photoevaporation dominates $\dot M_{\rm pho}$,  the disk dispersal timescale $\taud$,  and the inner disk cavity expansion speed $\vr$.

\subsection{Planet-gas disk interaction.} 

Here we summarize the formulas of planet migration torques used in our study based on Ref.\cite{Liu2017a}.
When the planet is far from the inner edge of the disk, it feels the torque from disk gas on both sides.
The two-sided torque $ \Gamma_\mathrm{2s} $ is adopted from Eq. (49) of Ref.\cite{Paardekooper2010}: 
\begin{equation}
     \frac{\Gamma_\mathrm{2s}} {m_{\rm p} (r_{\rm p} \Omegap)^2} =   -2.3  \qd \frac{\qp}{h^2},
    \label{eq:torq_2s}
\end{equation}
where $\Omegak$ is the Keplerian angular velocity, $r_{\rm p}$ is the distance from the planet to the central star, and $\qd {\equiv} \Sigmap r_{\rm p}^2 / M_{\star}$ and $\qp {\equiv} m_{\rm p}/M_{\star}$ are the mass ratio between the local gas disk and star and mass ratio between the planet and star, respectively.
When the planet is at the disk edge, only the one-sided torque $ \Gamma_\mathrm{1s} $ exists: 
 \begin{equation}
\frac{\Gamma_\mathrm{1s}} {m_{\rm p} (r_{\rm p} \Omegap)^2} = C_\mathrm{hs} \qd  \left(\frac{\qp}{h^{3}}\right)^{1/2} + C_{\rm L} \qd \frac{\qp}{h^3}, 
 \label{eq:torq_1s}
\end{equation}
where the first and second terms on the right side of \eq{torq_1s} are the corotation and Lindblad torque components with  $C_\mathrm{hs}{=}2.46$ and $C_\mathrm{L}{=}{-}0.65$.  In the above equations, the corotation torques are expressed for planets on circular orbits. The saturation of the corotation torque due to non-zero eccentricity is accounted for by adopting Eq.(13) of Ref.\cite{Liu2017a}.
To derive the corotation torque in \eq{torq_1s}, we assume that at the disk inner edge, the gas removal time $t_{\rm removal}$ is faster than the gas libration time in the planet horseshoe region $t_{\rm lib}$.
We show that this is a justified treatment as follows.
First, at the inner disk edge the gas removal time should be no longer than the viscous diffusion time $t_{\rm vis}$; otherwise the gas would accumulate.
Thus, we have $t_{\rm removal}{\lesssim}t_{\rm vis} {\simeq} x_{\rm hs}^2/\nu$, where $x_{\rm hs} {\sim} r_{\rm p} \sqrt{\qp/h}$ is the half-width of the planet horseshoe region.
The gas libration time for a planet can be written as  $t_{\rm lib} {\sim} 8 \pi r/(3 \Omega_{\rm K} x_{\rm hs}) $.
Then $t_{\rm removal} {<} t_{\rm lib}$ is required for the one-sided corotation torque in \eq{torq_1s}.
The above condition is satisfied as long as $t_{\rm vis}{<}t_{\rm lib}$, and one can obtain that $q {\lesssim} (8\pi /3 \alpha)^{2/3} h^{7/3}{\simeq}2.4 {\times} 10^{-4}$ by adopting $\alpha{=} 0.005$ and $h {=}0.07$ (at the location of the $3$:$2$ MMR with Jupiter).
In other words, in order to fulfill the condition of large amplitude one-sided corotation torque, the planet needs to be no more massive than Saturn in our disk model.

For a general situation, the total torque can be expressed by interpolating the torques between the two regimes:
 \begin{equation}
\Gamma = f \Gamma_{\rm 1s} + (1-f) \Gamma_{\rm 2s}, 
 \label{eq:torq}
\end{equation}
where $f {= } \exp \left[-(r_{\rm p}-r_{\rm in} )/x_{\rm hs}\right]$ is a smooth fitting function and $f{=}1$ ($f{=}0$) refers to the planet at the disk edge  (far away from the disk edge).  
The disk torque is added into the the equation of the planet motion in a cylindrical coordinate\cite{Liu2015}: 
  \begin{equation}
 \frac{d v_{\theta}}{dt} =  \frac{\Gamma}{m_{\rm p} r_{\rm p}},
 \end{equation}
  \begin{equation}
  \frac{d v_{\rm r}}{dt} = - \frac{v_{\rm r}}{t_{\rm ecc}},
 \end{equation}
  \begin{equation}
  \frac{d v_{\rm  z}}{dt} = - \frac{v_{\rm z}}{t_{\rm inc}}.
 \end{equation}
 Both orbital migration and eccentricity/inclination damping are included (see Eqs.22-23 of Ref.\cite{Liu2015}), where $t_{\rm ecc}{=}t_{\rm inc}$ is assumed.
Although the above equations are derived analytically, we note that the termination of planet migration at the inner disk edge due to one-sided torques 
is also obtained in hydrodynamic simulations\cite{Masset2006, Romanova2019}.
 
In order for planet outward migration with the retreating disk, the one-sided corotation torque needs to be larger than the one-sided Lindblad torque.
Thus, from \eq{torq_1s} the planet needs to satisfy the condition: $\qp {<}(C_{\rm hs}/C_{\rm L})^2 h^3$.
The above torque formulas are derived in the linear regime. However, when the planet is massive to clear the local surrounding gas in its horseshoe region\cite{Lin1986}, the disk feedback is non-trivial and \eqs{torq_2s}{torq_1s} are not applicable anymore.
Because the positive corotation torque diminishes due to gap formation, the planet beyond the gap-opening mass fails to undergo outward migration.
The gap-opening requires the planet's Hill sphere $R_{\rm H} {\equiv} (m_{\rm p}/3M_{\star})^{1/3}r$ to be larger than the disk scale height $H$.
Therefore,  $\qp {< }3h^3$ is needed for torques in the linear regime.
Besides, dedicated hydrodynamic simulations indicated that the planet needs to fulfill this criterion as well\cite{Crida2006}:
\begin{equation}
\frac{3}{4} \frac{H}{R_{\rm H}} + \frac{50}{q_{\rm p} R_{\rm e}}  >1, 
 \label{eq:gap1}
\end{equation}
where the Reynolds number $R_{\rm e}{=}r^2\Omegak/\nu$.
We name $q_{\rm p,c}$ as the maximum non-gap opening mass ratio obtained from \eq{gap1}.
To summarize, the planet-to-star mass ratio needs to satisfy the condition:
\begin{equation}
 q_{\rm p} {<} q_{\rm gap}{=} \min[(C_{\rm hs}/C_{\rm L})^2 h^3, 3h^3, q_{\rm p,c}]
  \label{eq:gap2}
\end{equation} for planet outward migration at the inner disk edge. 

For our fiducial disk model, Saturn, Uranus, and Neptune are in the linear type I torque regime.
More massive Jupiter is in the type II gap-opening regime, and rebound fails to operate for Jupiter.
Since the timescale of type II migration is much longer than that of the type I, and our study merely focuses on the time associated with rapid disk dispersal, we neglect the migration of Jupiter for simplicity. 
We assume that the giant planets have reached their present-day masses before the onset of final disk dispersal, so the above assessment of gap-opening is based on their full masses.
Nonetheless, our model holds as long as proto-Jupiter has completed its main gas accretion and became more massive than Saturn before rebound operates.

We note that the above migration condition requires a disk with moderately high viscosity and aspect ratio. The simulations are performed with this fiducial disk setup unless otherwise stated.  
A too low $\alpha$ or $h$ will also cause Saturn to open a deep gap.
Then, the picture changes since both Jupiter and Saturn undergo slow type II migration. 
We also conduct a subset of simulations in a disk with low viscosity and aspect ratio.  The influence of these two parameters is investigated in \textbf{low-viscosity disks} section. 

\subsection{Planet-planetesimal disk interaction.}
The outer planetesimal disk exchanges angular momentum with the giant planets, resulting in the expansion of the planet's orbits with damped eccentricities and inclinations\cite{Fernandez1984}. Such a planet-planetesimal disk interaction is often considered to be a  trigger for the late giant planet instability, which played a crucial role in shaping the final architecture of the Solar System\cite{Tsiganis2005},  typically taking place a few hundreds of Myr after Solar System formation\cite{Gomes2005}.
However, the orbits of the fully-formed inner terrestrial planets are likely destabilized by such a late instability\cite{Agnor2012, Kaib2016}, motivating the consideration of an earlier instability\cite{Clement2018}.

During the gas-rich disk phase, a planetesimal of a radius $R_{\rm plt}$ experiences aerodynamic gas drag, and their orbital decay timescale can be expressed as: 
  \begin{equation}
t_{\rm drag}  \simeq 1.2 \times 10^{8}  {\rm yr} \left(  \frac{R_{\rm plt}}{10 \ \rm km}  \right) \left(  \frac{\Sigma_0}{10^{3} \ \rm g \ cm^{-2}}  \right)^{-1}  \left(  \frac{h_{\rm 0}}{0.04}  \right)^{-1}  \left(  \frac{\rho_{\bullet}}{1.5 \ \rm g \ cm^{-3}}  \right)  \left(  \frac{a}{30 \rm \  AU}  \right)^{9/4},   
 \label{eq:drag}
\end{equation}
where $\Sigma_0$ and $h_0$ are the gas surface density at $1$ AU, and $\rho_{\bullet}$ is the internal density of the planetesimal.
We find that since $t_{\rm drag}$ is much longer than the gas disk lifetime, planetesimals in the proto-Kuiper Belt with radii larger than $10$ km experience negligible radial drift during the gas disk phase.

\subsection{Numerical methods.}
We perform numerical simulations using a modified version of the publicly available N-body code HERMIT4\cite{Aarseth2003} to study the evolution of multi-planet systems during gas disk dispersal.
The code includes the planet-gas disk interaction by implementing the previously mentioned torque recipes\cite{Liu2015, Liu2017a}. 
Besides, we run extended simulations to study the effect of a planetesimal disk on giant planets' orbital evolution in a gas-free environment. These simulations are conducted separately using the open-source N-body code MERCURY with a hybrid symplectic and Bulirsch-Stoer integator\cite{Chambers1999}.

\subsection{Gas disk study.}
We investigate the evolution of planetary systems during the final gas disk dispersal phase in a statistical manner. The initial disk and planet conditions are listed in Extended Data \tb{IC}.  
We consider three different planet configurations: all planets in nearly $2$:$1$ resonances, $3$:$2$ resonances, and a combination of $2$:$1$ and $3$:$2$ resonances.
Different initial numbers of planets have also been explored: $N{=}4$, $5$, and $6$. For each planetary configuration, we have performed $1000$ simulations by Monte Carlo sampling the disk properties ($\dot M_{\rm pho}$, $\taud$, and  $\vr$).
Importantly, we have considered both simulations with and without rebound to evaluate the efficacy of this mechanism.
Planets feel the classic Type I torques as in \eq{torq_2s} when rebound is absent, whereas they feel torques including the one-sided components as in \eqs{torq_1s}{torq} when rebound is present.

We start the initial planet period ratios $5\%$ higher than the exact resonant states.
To further set up the initial conditions, we integrate the planets for $0.5$ Myr with only the migration of the outermost planet turned on, and the gas surface density unchanged.
This ensured that the planets got into the desired resonant chains.
The eccentricities and inclinations of the planets are assumed to follow Rayleigh distributions, where the scale parameters $e_0 {=} 2 i_{0} {=} 10^{-3}$.
The other orbital phase angles are randomly selected between $0^{\circ}$ and $360^{\circ}$.
After the initial $0.5$ Myr integration, we turned on the migration for all planets, and the disk starts to deplete according to the sampled disk properties.
All parameter study simulations are terminated at $10$ Myr where gas disks are fully dissipated. 
 
Our goal is to demonstrate our new instability trigger, so we used two broad dynamical indicators to show that our simulated systems are indeed consistent with the global properties of the Solar System without attempting to match each detailed constraint.
The first is the normalized angular momentum deficit\cite{Laskar1997}: $ {\rm AMD} {=}  \sum {m_{\rm j} \sqrt{a_{\rm j}}} \left( 1 - \cos\left(i_{\rm j}\right)\sqrt{1-e_{\rm j}^2}\right) / \sum {m_{\rm j} \sqrt{a_{\rm j}}},$ where $m_{\rm j}$, $a_{\rm j}$, $e_{\rm j}$, and $i_{\rm j}$ are the mass, semimajor axis, eccentricity, and inclination of each giant planet.
The AMD increases with increasing orbital eccentricities and inclinations, because eccentric or inclined orbits have a lower $\vec{h}$-projected angular momentum than circular, co-planar ones at the same semimajor axes.
The second indicator is the radial mass concentration\cite{Chambers2001}:  ${\rm  RMC} =\max \left( \sum m_{j} / \sum m_{j }[\log_{10}(a/a_{j} )]^2 \right)$, which measures the degree of radial mass concentration in a given system, with higher RMC corresponding to a more tightly packed system.

 Compared to \Fg{5planet}, We show the effect of rebound on different initial potential giant planet architectures in Extended Data Figures \ref{fig:appendix_dynamics4planets}, \ref{fig:appendix_dynamics5planets} and \ref{fig:appendix_dynamics6planets}.
The model parameter setups and numerical outcomes can be found in Extended Data \tb{IC}.  In addition, the AMDs of systems with initially four, five, and six planets as a function of disk parameters are presented in Extended Data Figures \ref{fig:appendix_diskparameter4planets} and \ref{fig:appendix_diskparameter5planets}.
For instance, the upper panel of Extended Data Figure \ref{fig:appendix_diskparameter4planets}, provides the outcome of simulations where there were initially four giant planets in a chain of $2$:$1$ resonances (\textit{run\_A4R} in Extended Data \tb{IC}). 
For this setup, the Solar System analogs (black dots) are likely to form when $\dot M_{\rm pho}$ is lower than $10^{-9} \Msunyr$.

\subsection{Planetesimal disk study.}
In addition to the above simulations only considering a gas disk, we also ran simulations to account for an outer planetesimal disk.
Such a disk is expected to continuously exchange angular momentum with the giant planets on a much longer timescale, motivating our inclusion of a planetesimal disk only after the gas disk is entirely depleted.
We performed new sets of extended gas-free simulations for another $100$ Myr using the MERCURY code, where the initial orbital information of giant planets is adopted from the previous Hermite simulations at $t{=}10$ Myr.
For the purpose of illustration, we only perform a limited number of such extended simulations rather than extensive explorations in a multi-parameter space.

We adopted the test particle approach to reduce the computational cost.
The disk contains $1000$ test particles and each particle represents a swarm of real planetesimals at similar positions and velocities.
The particles feel the gravitational forces from the planets, but the interactions between themselves are neglected. 
We assumed that the planetesimal disk extends from $20$ to $30$ AU, with a surface density profile of  $\Sigma_{\rm plt} \propto r^{-1}$.
The total disk mass is varied from $5$, $10$ and $20 \Me$.
These test particles are initialized on nearly circular and coplanar orbits, and their eccentricities and inclinations follow Rayleigh distributions with $e_0 {=}2 i_{0} {=} 10^{-3}$.
We also tested for numerical convergence using a total number of $500$ test particles.
The results are found to be consistent between the explored resolutions.

\subsection{Secular mode analysis}

Here we check the secular dynamics and eccentricity mode of the resulting planetary systems.  Since Jupiter and Saturn are the dominant mass contributors among Solar System planets, the secular dynamics of the system can be approximately traced by solving the Lagrange-Laplace equations for the Jupiter-Saturn pair\cite{Morbidelli2009}.
The eccentricities and longitudes of perihelia of Jupiter and Saturn can be written as: 
\begin{align}
\begin{split}
e_{\rm J} \cos \overline{\omega}_{\rm J} &=M_{55} \cos(g_5 + \beta_5) + M_{56} \cos(g_6 + \beta_6), \\
e_{\rm J} \sin \overline{\omega}_{\rm J} &=M_{55} \sin(g_5 + \beta_5) + M_{56} \sin(g_6 + \beta_6), \\
e_{\rm S} \cos \overline{\omega}_{\rm S} &=M_{65} \cos(g_5 + \beta_6) + M_{66} \cos(g_6 + \beta_6), \\
e_{\rm S} \sin \overline{\omega}_{\rm S} &=M_{65} \sin(g_5 + \beta_6) + M_{66} \sin(g_6 + \beta_6), 
\label{eq:JSmode}
\end{split}
\end{align}
where $g_{\rm i}$ are the eigenfrequencies for the precession of perihelia, $\beta_{\rm i}$ are the phase angles, and $ M_{\rm jk}$ are the coefficients of the corresponding eccentricity amplitudes.
The subscript $i$ refers to the eight Solar System planets, and  $5$ and $6$ represent Jupiter and Saturn.
We note that the perturbations from the other planets are relatively small compared to these from Jupiter and Saturn.
Thus, these four amplitudes and two eigenfrequencies can be treated as a good approximation for the secular evolution of Jupiter and Saturn.

A proper excitation of Jupiter's eccentricity mode, particularly $M_{55}$ to $0.044$, is the most difficult property of the gas giant planets' secular architecture to match; previous numerical simulations such as those of the Nice model or other alternatives generally yield lower values\cite{Morbidelli2009, Nesvorny2012, Clement2021}. 
Extended Data Figure \ref{fig:emode} shows the amplitude of the $M_{55}$ mode for the simulated planetary systems lies between $0.005$ and $0.05$.
Simulations that only consider evolution in the depleting gas disk generate higher amplitudes compared to those that additionally account for the planetesimal disk.
This is simply because the dynamical friction from the outer planetesimal disk continuously damps the eccentricities of the inner giant planets.

\subsection{Low-viscosity disks}
In order to address whether the rebound mechanism can trigger a dynamical instability in low-viscosity disks where Saturn carves a deep gap, we conducted a new set of simulations focusing on the interactions with the gas disk, i.e. no planetesimal disk included.
We adopted a layered accretion disk model\cite{Liu2019}, where $\alpha$ represents the averaged global disk angular momentum transport efficiency and $\alpha_{\rm t}$ corresponds to the local turbulent viscosity strength at the disk midplane.
Therefore, $\alpha $ sets the gas surface density while the gap opening and planet migration are governed by $\alpha_{\rm t}$.
The two values are set to be equal in the fiducial disk model ($\alpha{=}\alpha_{\rm t}{=}0.005$).
Here, we also assume $\alpha{=}0.005$ but vary $\alpha_{\rm t}$ to test its influence on the system's instability rate.  
We focused on the initial $3$:$2$ resonant configurations of five and four planets (the same as $\textit{run\_B5R}$ and $\textit{run\_B4R}$), and we specified the disk parameter distributions such that $\dot M_{\rm pho}$, $\taud$, and $v_{\rm r}$ are log-uniformed selected from [$10^{-9.5},10^{-9}$] $ \Msyr$,  [$10^{5.5},10^{6} $] yr, and [$40,120$] AU/Myr, respectively.
The adopted midplane $\alpha_{\rm t}$ and the disk aspect ratio at the onset of the inner disk edge $h_{\rm in}$ are illustrated in Extended Data \Fg{gap}.  
 
In a low viscosity disk environment, the outward-sweeping disk edge only directly affects the orbits of the ice giant planets.
Both Jupiter and Saturn are in the gap-opening regime, and their slow type II migration is neglected during the rapid gas disk dispersal phase. 
Although Jupiter and Saturn cannot move out of resonance directly as the disk edge sweeps by, the subsequent instabilities triggered among the ice giants have a chance to propagate to the two larger gas giant planets (see an example of Extended Data \Fg{lowalpha}).

As shown in Extended Data \tb{ins}, orbital restructuring and system-wide instability are still common outcomes.
A high fraction of systems experienced instabilities during the first $10$ Myr, most of which take place within $1{-}3$ Myr simultaneously with the rapid disk dispersal (Extended Data \Fg{instability}a).
In most circumstances, dynamical instabilities start among the ice giant planets and rapidly propagate to destabilize Jupiter and Saturn's orbits (Extended Data \Fg{instability}b).
The instability propagation rate is higher than $90\%$ in general.
This is true for both four and five planet initial configurations, which means that the propagation of the instability from the ice giants to Jupiter and Saturn does not rely on the presence of an additional ice giant. 
Through the above exploration, we conclude that the rebound mechanism is a viable instability trigger even in low-viscosity disks, as long as at least one planet is below the gap-opening threshold.

\end{methods}


\newpage

\setcounter{figure}{0} 

\begin{figure}
    \includegraphics[width=16cm]{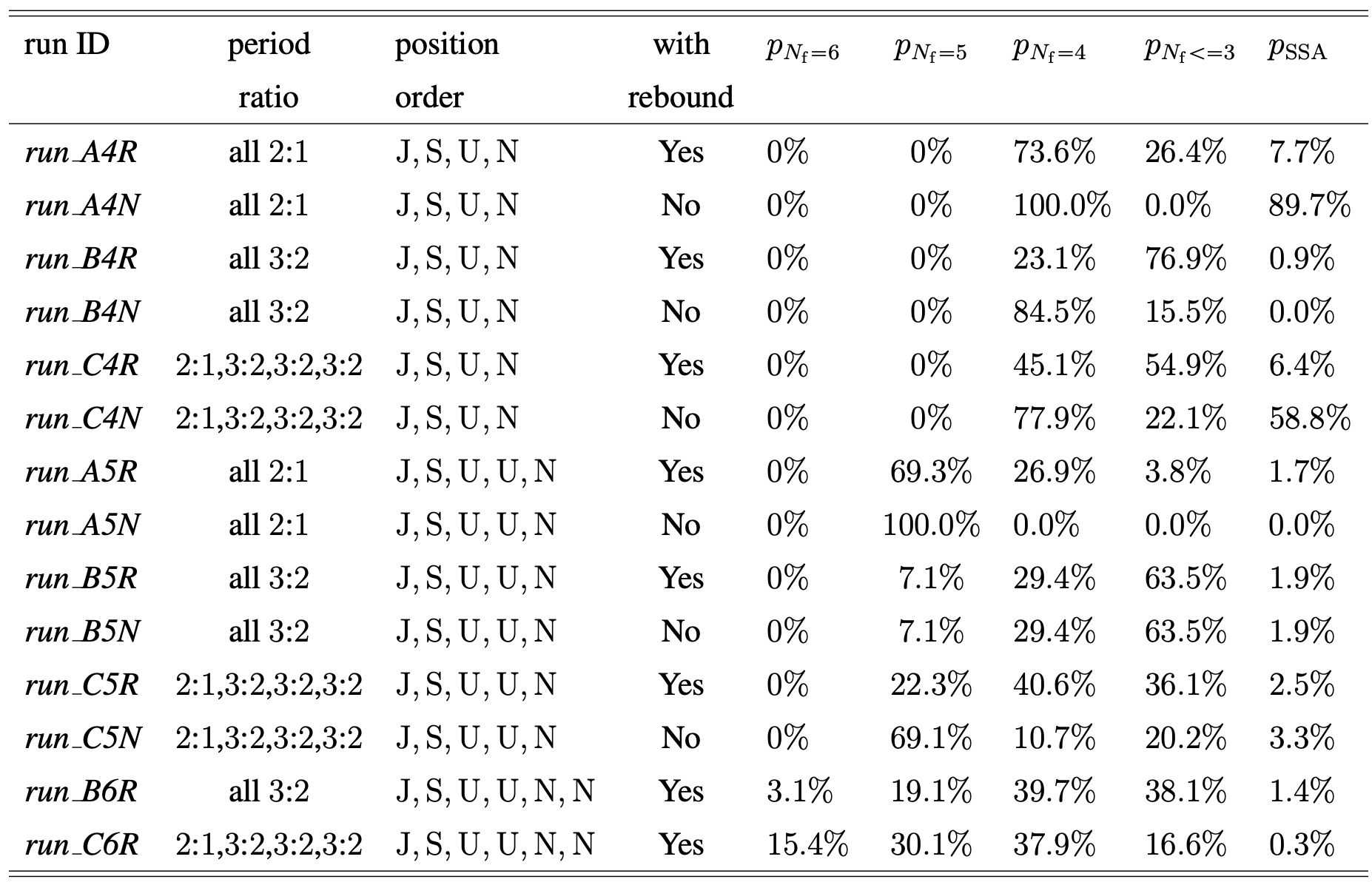}
    \renewcommand\figurename{Extended Data Table }
    \caption{\textbf{
  Initial conditions and statistical outcomes of the gas disk parameter study.
    }
    The first to fourth columns correspond to the name of the run,  planet period ratio,  position order (from inner to outer) and the option of simulations including rebound or not, where J, S, U, N are short for the planet with the mass of Jupiter, Saturn, Uranus, and Neptune, respectively.   The disk parameters $\dot M_{\rm pho}$, $\taud$, and $v_{\rm r}$ are log-uniformed selected from [$10^{-9.5},10^{-8.5}$] $ \Msyr$,  [$10^{5},10^{6} $] yr, and [$20,200$] AU/Myr,  respectively.  The fifth to eighth columns show the probability of systems with a final number of planets $N_{\rm f}{=}6$, $5$, $4$, and ${\leq}3$, respectively.
   The ninth column represents the probability of forming Solar System analogs, defined that systems survive with four planets in the right position order and their AMDs and RMCs are within a factor of three compared to the Solar System. The orbits of the current giant planets are adopted from Table 1 of Ref.\cite{Nesvorny2012}.
    }
    \label{tab:IC}
\end{figure}

\begin{figure}
    \includegraphics[width=12cm]{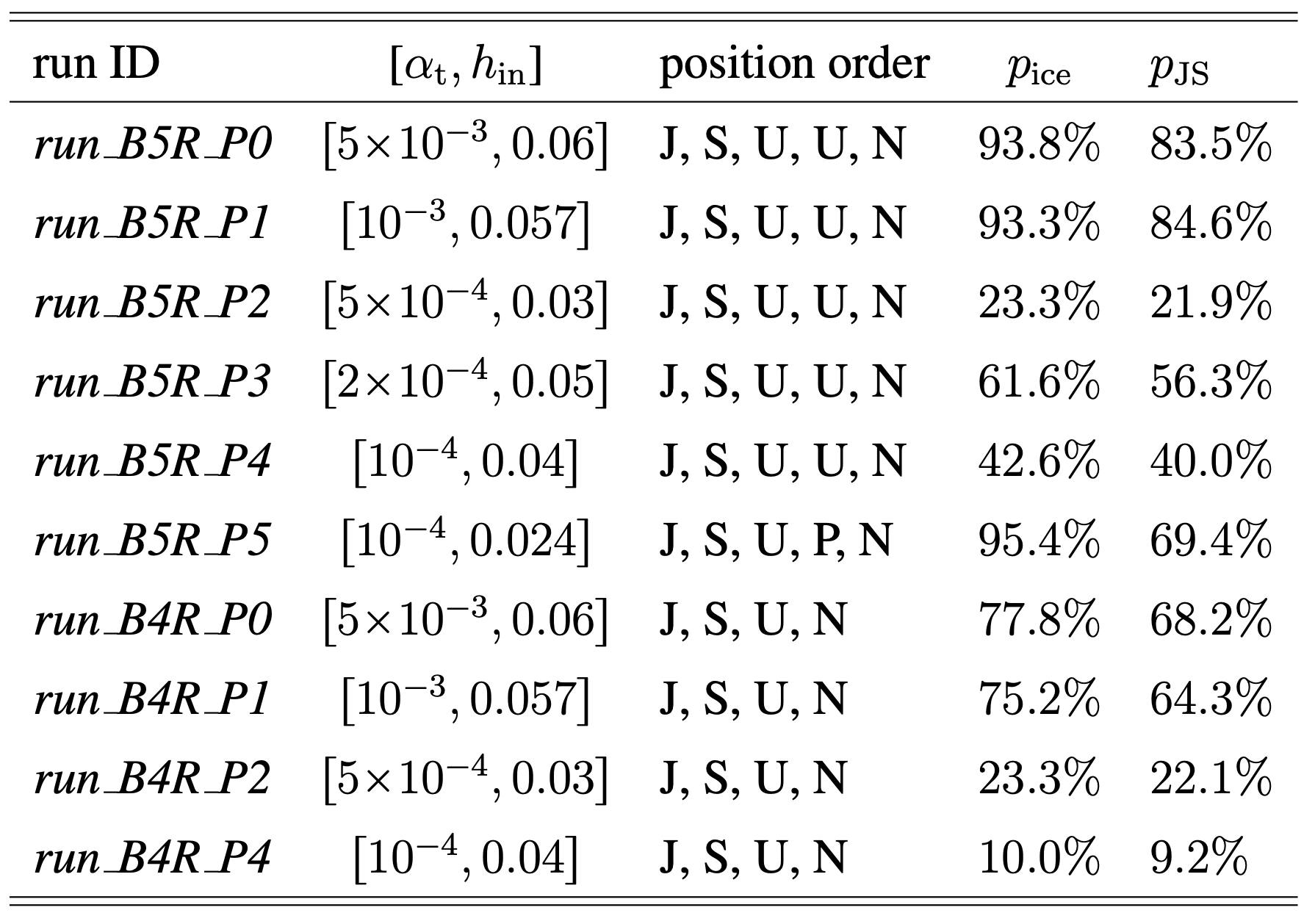}
    \renewcommand\figurename{Extended Data Table}
    \caption{\textbf{
    Instability statistics including low-viscosity disks.
    }
    The first and second columns provide the name of the run and the adopted $\alpha_{\rm t}$ and $h_{\rm in}$.   The third column lists the planet position order,  where in $\textit{run\_B5R\_P5}$ the mass of one ice giant is chosen to be $8 \Me$. The instability probability is given by $p_{\rm ice/JS}$, where the subscript $\rm ice$ or $\rm JS$ represents that the dynamical instability occurs for ice giant planets or the instability spreads to the Jupiter and Saturn pair.
    }
    \label{tab:ins}
\end{figure} 

\setcounter{figure}{0}  

\begin{figure}
    \includegraphics[width=14cm]{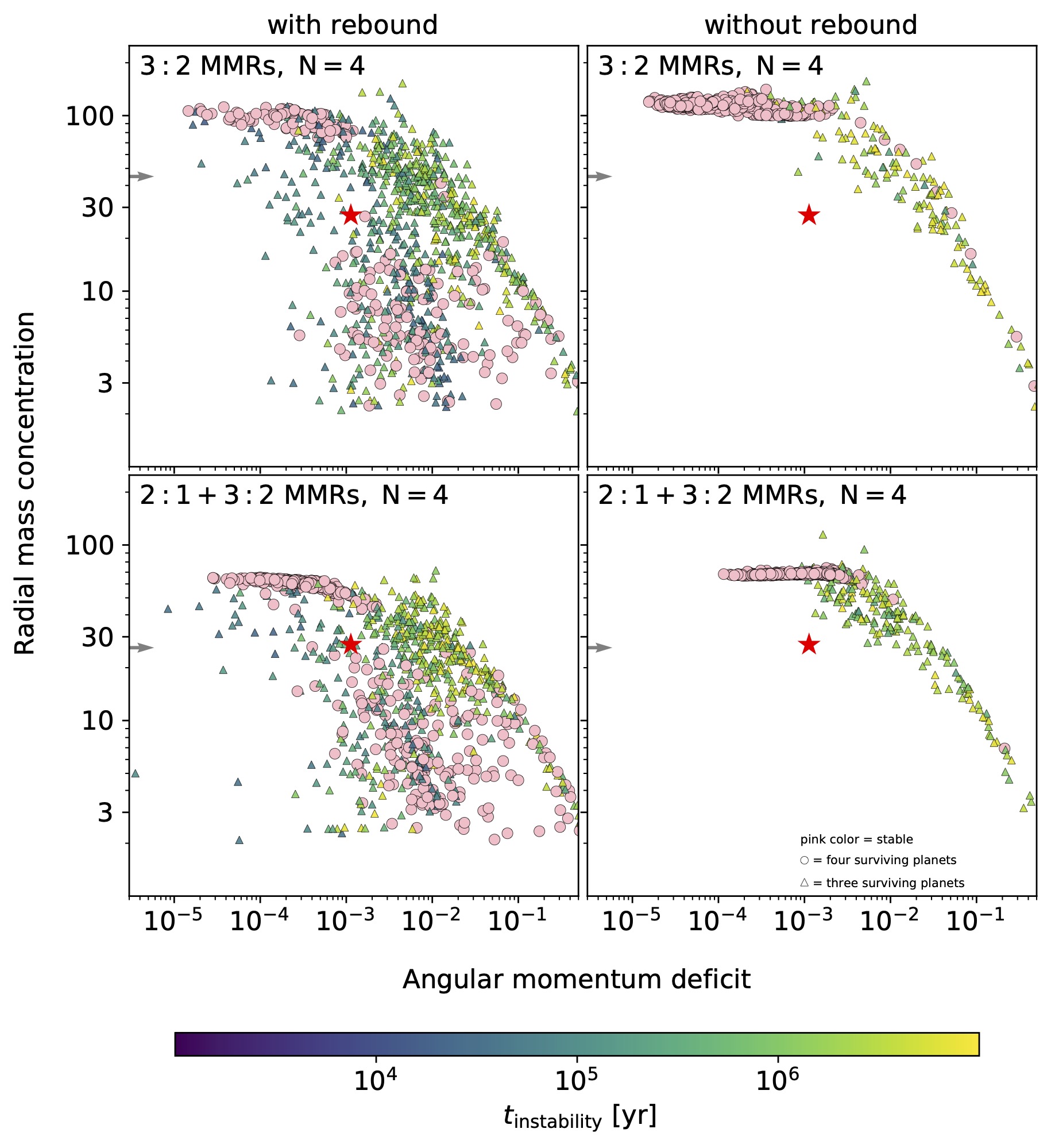}
    \renewcommand\figurename{Extended Data Figure}
    \caption{\textbf{
    Metrics for surviving planetary systems of a subsample of our simulations in matching the Solar System, comparable to \Fg{5planet}.
    }
    The simulations on the left included the rebound effect and those on the right did not.
    Each simulation started with our four present-day giant planets.
    In the top panels, the giant planets were initially placed in a chain of $3$:$2$ orbital resonances.
    In the bottom panels, Jupiter and Saturn were initially in a $2$:$1$ resonance and each other neighboring planet pair was in a $3$:$2$ resonance.
    Each symbol represents the outcome of a given simulation at $t{=}10$ Myr.
    The color indicates the timing of the instability after the start of gas disk dispersal; pink systems did not undergo an instability (no collision and/or ejection).
    Circles and triangles correspond to systems with four and three or fewer surviving planets, respectively.
    The arrow gives the initial radial mass concentration of the system.
    The Solar System is marked as a red star for comparison.
    }
    \label{fig:appendix_dynamics4planets}
\end{figure}

\begin{figure}
    \renewcommand\figurename{Extended Data Figure}
    \includegraphics[width=14 cm]{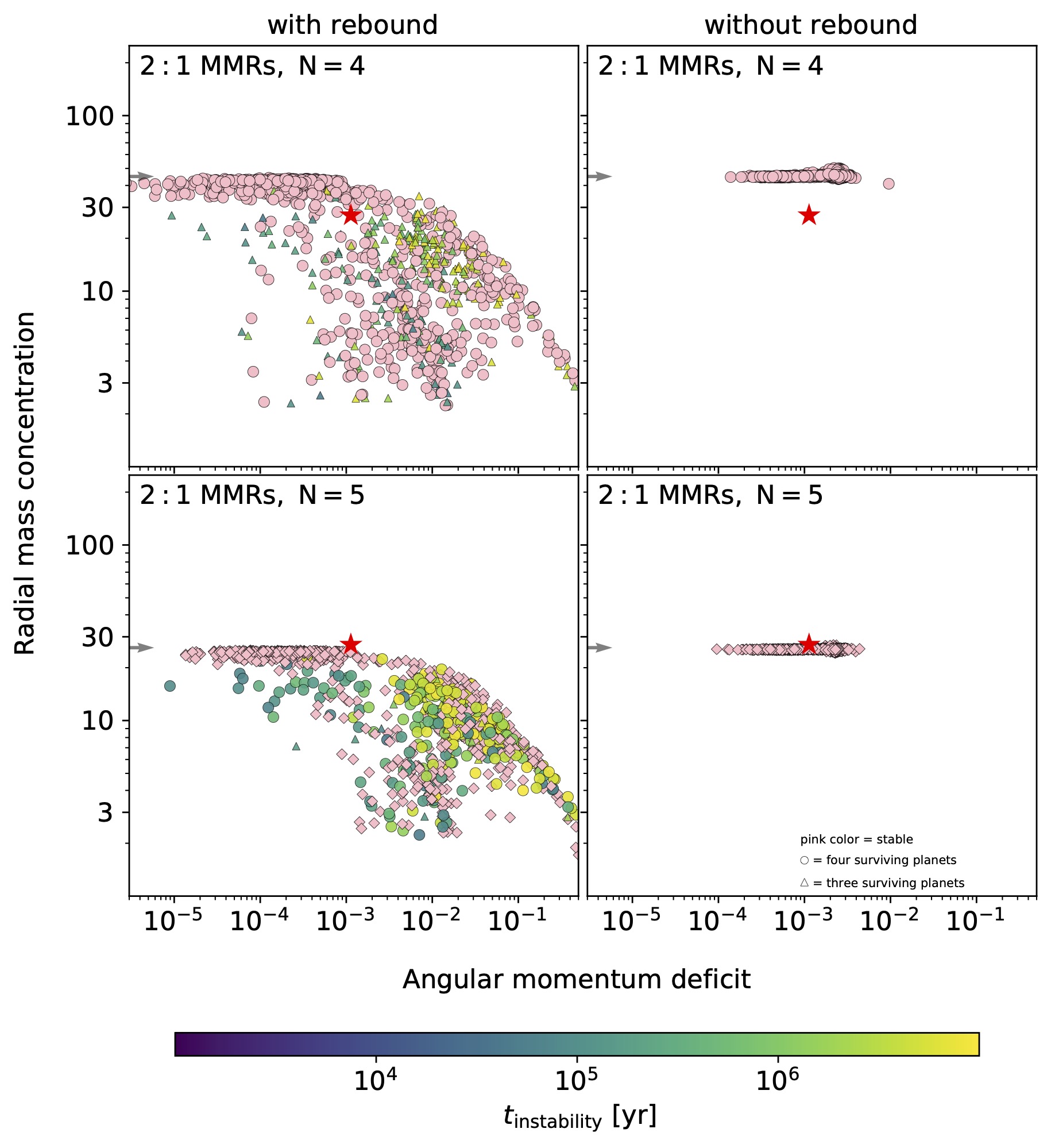}
    \caption{\textbf{
    Metrics for surviving planetary systems of a subsample of our simulations in matching the Solar System, comparable to \Fg{5planet}. 
    }
    The simulations on the left included the rebound effect and those on the right did not.
    Each simulation started with the giant planets in a chain of $2$:$1$ orbital resonances.
    In the top panels, we initially included only our four giant planets, but in the bottom panels, we added an additional ice giant at the start of the simulation.
    Each symbol represents the outcome of a given simulation at $t{=}10$ Myr.
    The color indicates the timing of the instability after the start of gas disk dispersal; pink systems did not undergo an instability (no collision and/or ejection).
    Diamonds, circles, and triangles correspond to systems with five, four, and three or fewer surviving planets, respectively.
    The arrow gives the initial radial mass concentration of the system.
    The Solar System is marked as a red star for comparison.}
    \label{fig:appendix_dynamics5planets}
\end{figure}

\begin{figure}
    \renewcommand\figurename{Extended Data Figure}
    \includegraphics[width=14 cm]{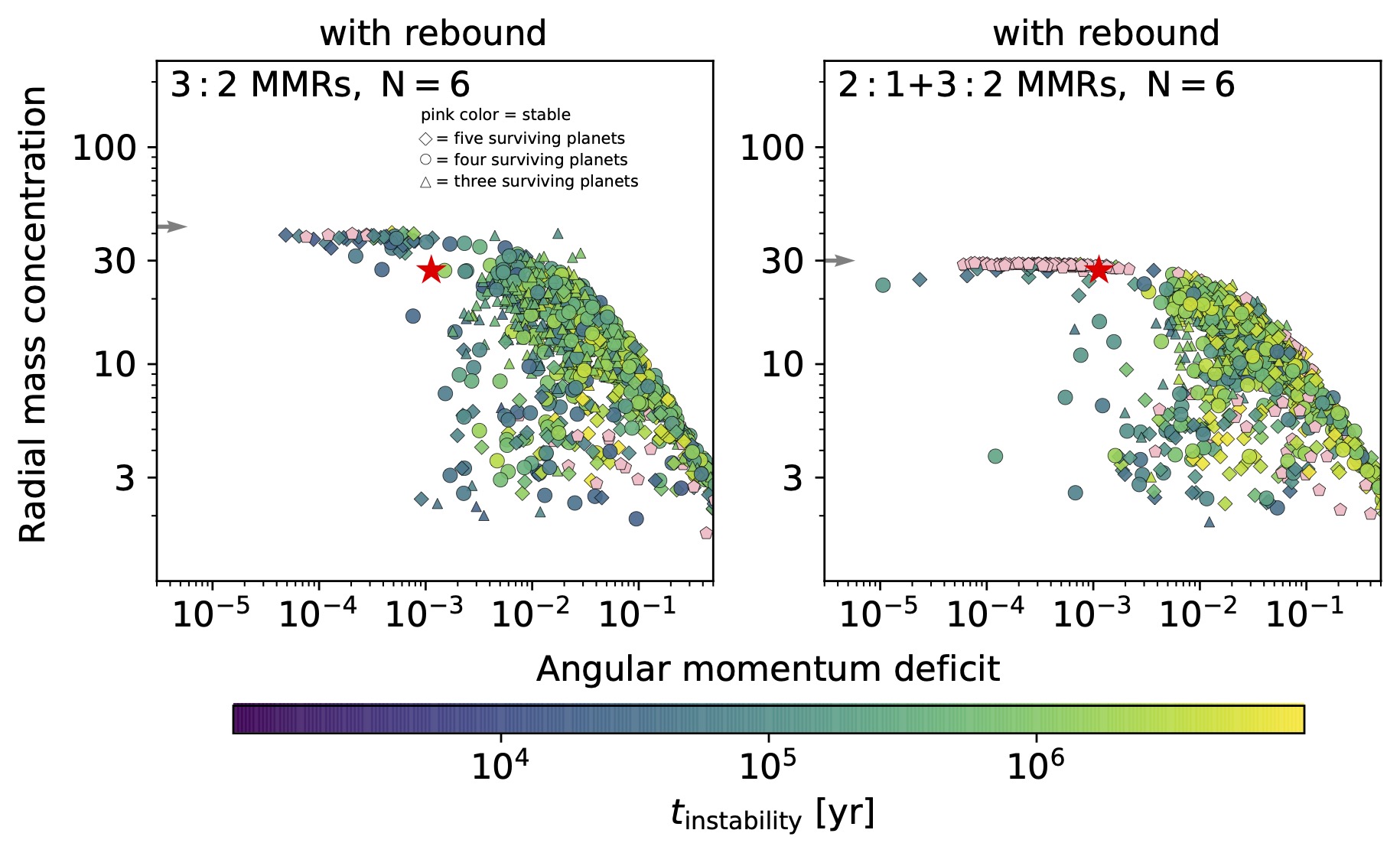} 
    \caption{\textbf{
    Metrics for surviving planetary systems of a subsample of our simulations in matching the Solar System, comparable to \Fg{5planet}.
    }
    Both panels, left and right, included the rebound effect.
    Each simulation started with our four present-day giant planets plus two additional ice giant planets.
    In the left panel, the giant planets are in a chain of $3$:$2$ orbital resonances. In the right panel, Jupiter and Saturn were initially in a $2$:$1$ resonance and each other neighboring planet pair was in a $3$:$2$ resonance.
    Each symbol represents the outcome of a given simulation at $t{=}10$ Myr.
    The color indicates the timing of the instability after the start of gas disk dispersal; pink systems did not undergo an instability (no collision and/or ejection).
    Pentagons, diamonds, circles, and triangles correspond to systems with six, five, four, and three or fewer surviving planets, respectively.
    The arrow gives the initial radial mass concentration of the system.
    The Solar System is marked as a red star for comparison.}
    \label{fig:appendix_dynamics6planets}
\end{figure}

\begin{figure}
    \renewcommand\figurename{Extended Data Figure}
    \includegraphics[width=16cm]{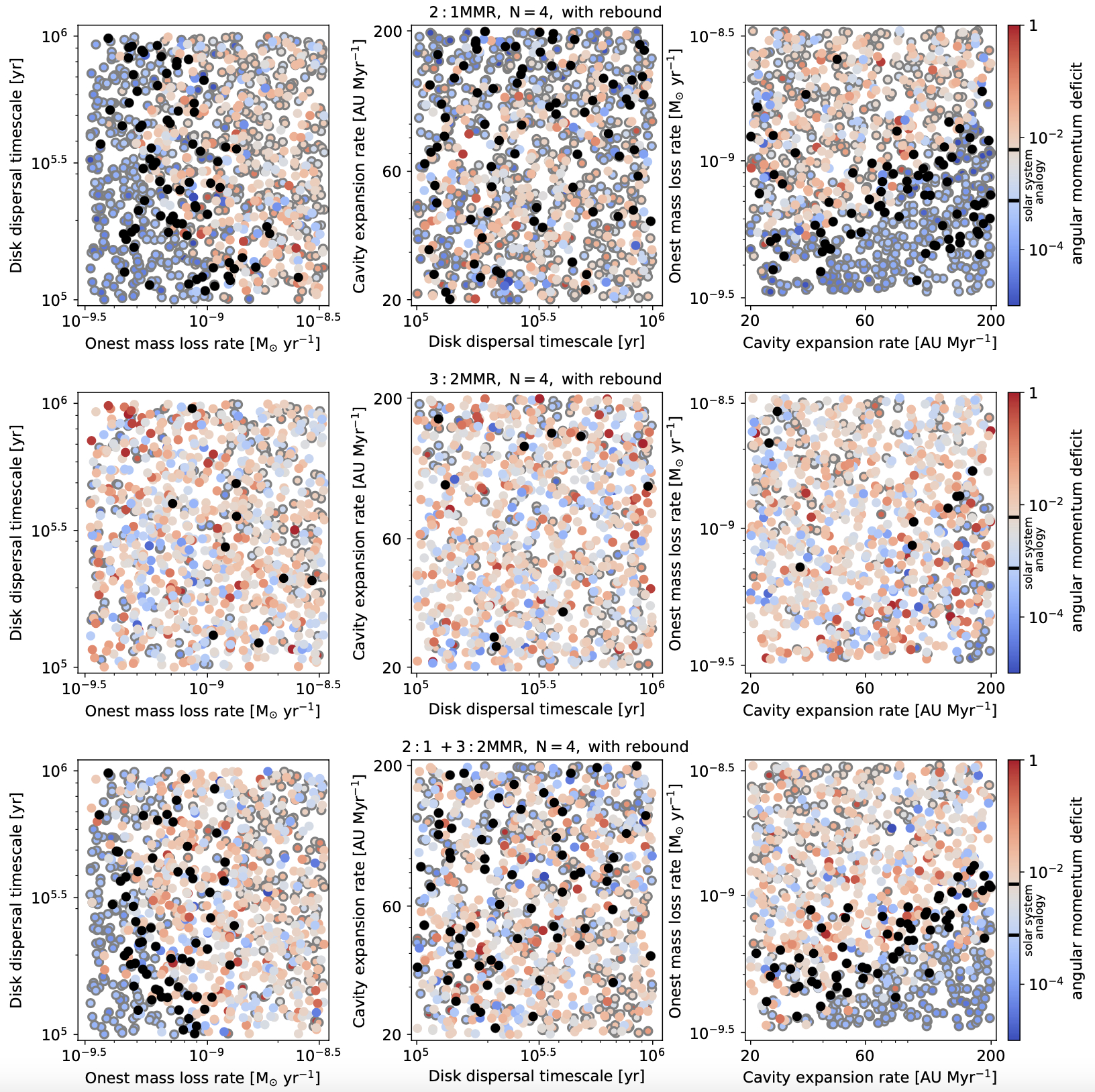}
    \caption{\textbf{Outcomes of rebound simulations with initially four planets as a function of disk parameters: onset mass-loss rate, the disk dispersal timescale, and the rate of expansion of the inner cavity.}
    Each simulation started with our four present-day giant planets in a chain of $2$:$1$ orbital resonances (top panels), a chain of $3$:$2$ orbital resonances (middle panels), or a combination of a $2$:$1$ orbital resonance and an ensuing chain of $3$:$2$ orbital resonances (bottom panels).
    The color bar corresponds to the system's angular momentum deficit (AMD).
    The circles with a grey edge color refer to the systems whose planets all survive in the end, while the black dots represent the Solar System analogs, defined as systems with four surviving planets in the correct order and their AMDs and RMCs are within a factor of three compared to those of our Solar System.}
    \label{fig:appendix_diskparameter4planets}
\end{figure}

\begin{figure}
    \renewcommand\figurename{Extended Data Figure}
    \includegraphics[width=12cm]{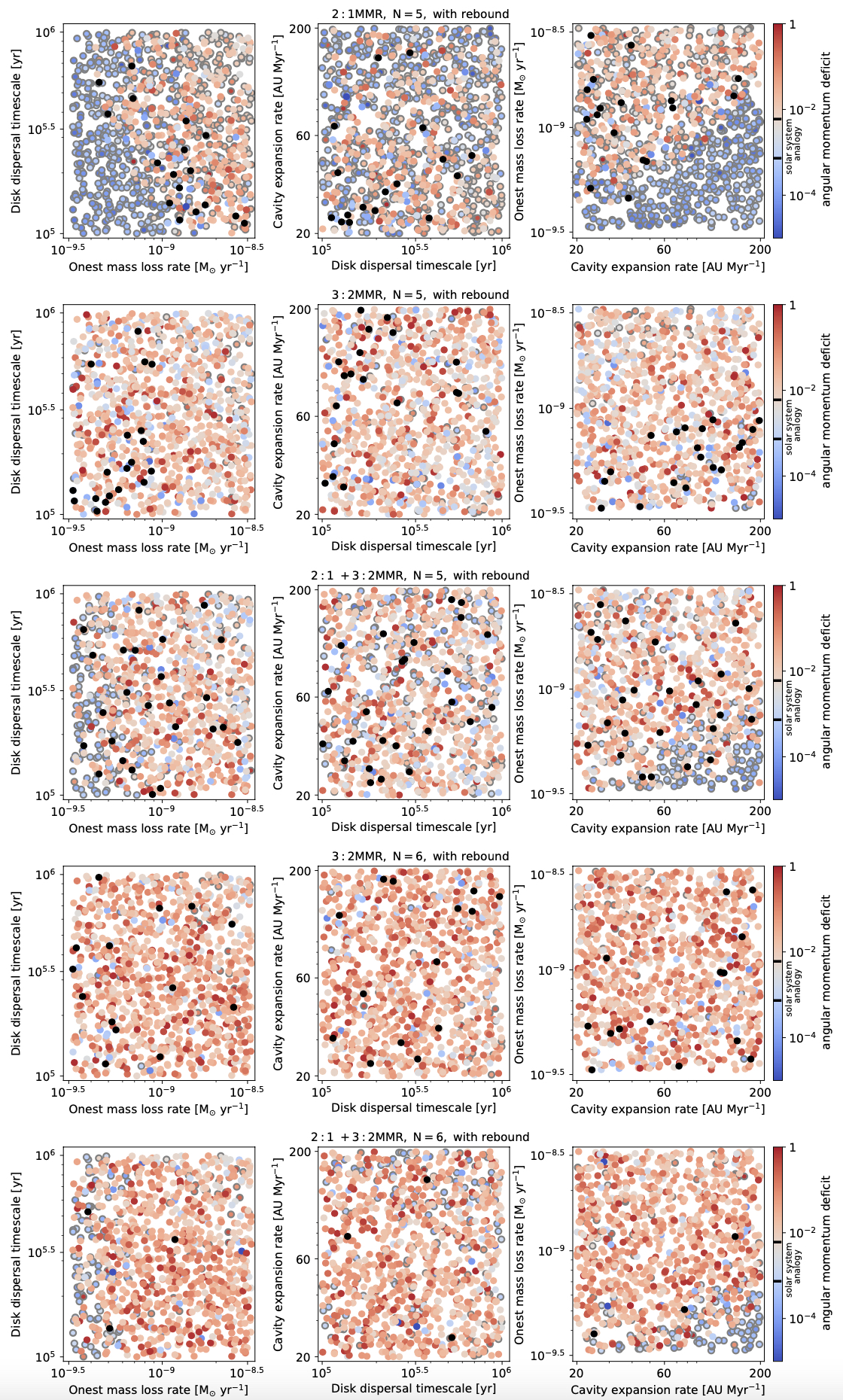}    
    \caption{\textbf{Outcomes of rebound simulations with initially five and six planets as a function of disk parameters: onset mass-loss rate, the disk dispersal timescale, and the rate of expansion of the inner cavity, comparable to Extended Data \Fg{appendix_diskparameter4planets}. } 
  Each simulation started with our four present-day giant planets plus one additional ice giant planet in a chain of $2$:$1$ resonances ($1$st row), a chain of $3$:$2$ resonances ($2$nd row), or a combination of  $2$:$1$ and  $3$:$2$ resonances ($3$rd row),  or started with our four present-day giant planets plus two additional ice giants in a chain of $3$:$2$ resonances ($4$th row), or a combination of $2$:$1$ and $3$:$2$ resonances ($5$th row).   }
    \label{fig:appendix_diskparameter5planets}
\end{figure}

\begin{figure}
    \renewcommand\figurename{Extended Data Figure}
    \includegraphics[width=16cm]{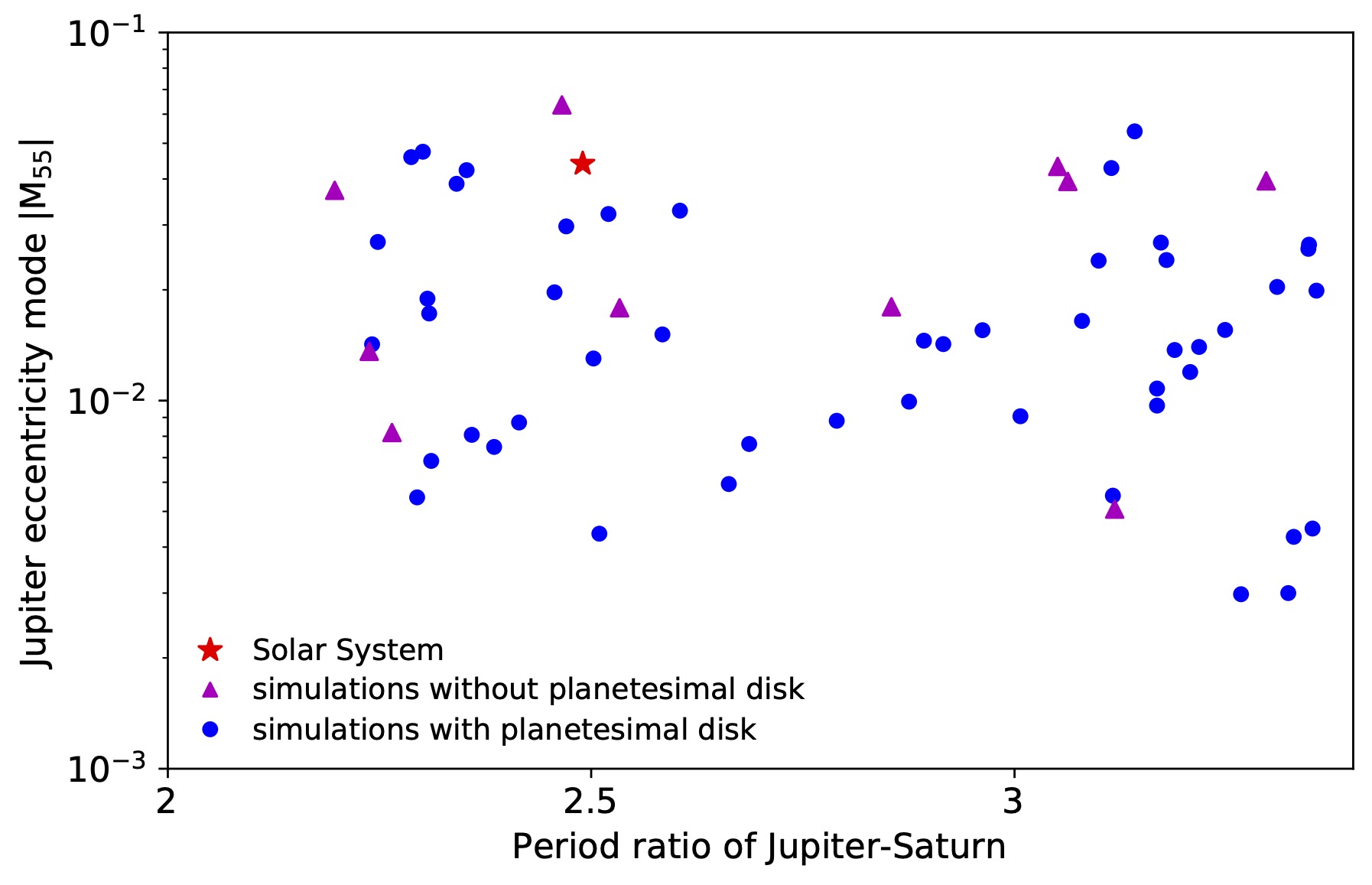}
    \caption{\textbf{Jupiter's eccentricity mode $M_{55}$ as a function of the period ratio of Saturn to Jupiter obtained in simulations with and without a planetesimal disk.}
    The simulations without and with planetesimal disks are plotted in triangles and circles, respectively, and the Solar System is marked as a star.
    Only systems that finish with four planets are shown here.}
    \label{fig:emode}
\end{figure}

\begin{figure}
    \renewcommand\figurename{Extended Data Figure}
    \includegraphics[width=16cm]{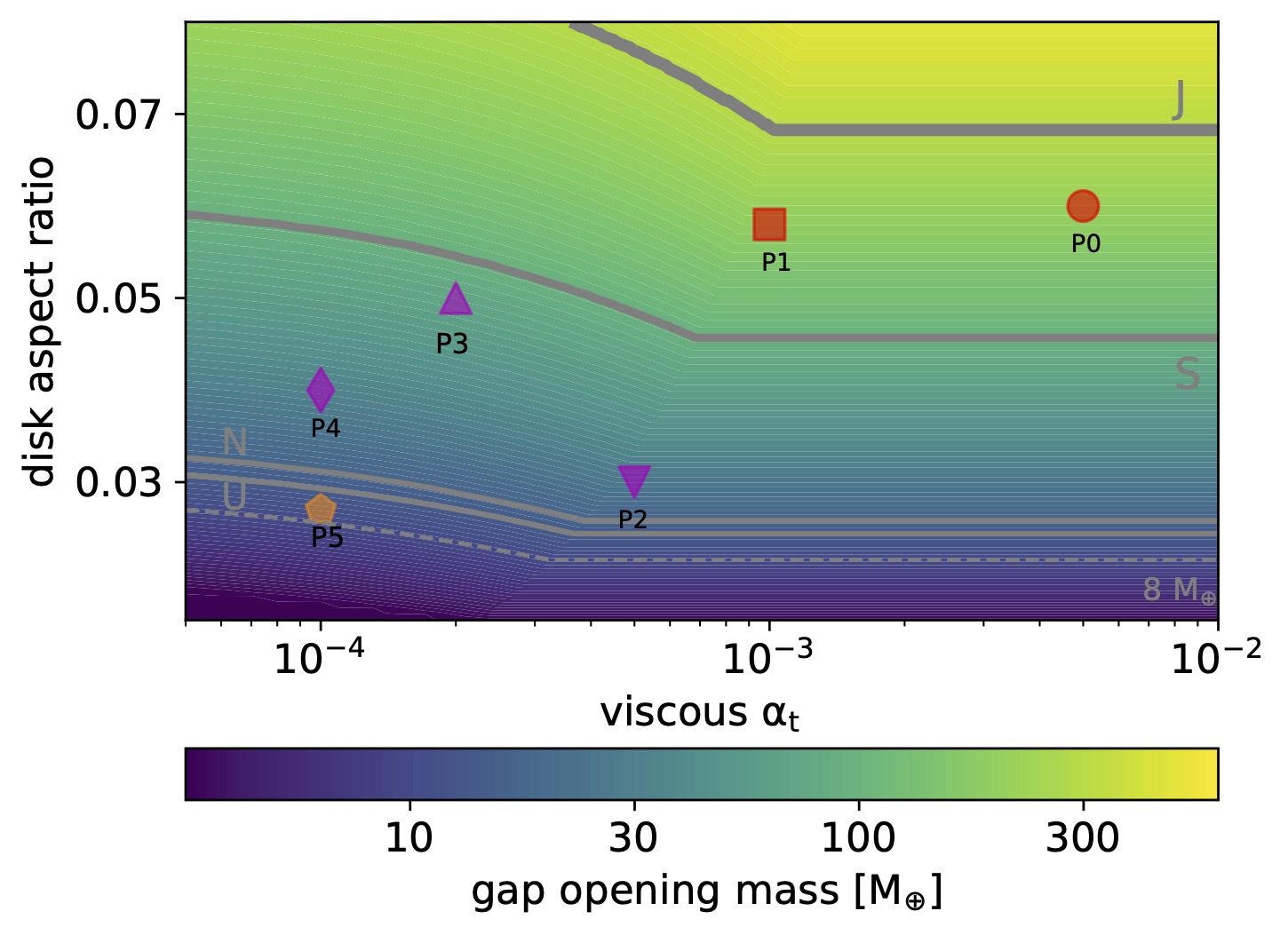} 
    \caption{\textbf{Gap opening mass as a  function of disk aspect ratio and midplane viscous $\alpha_{\rm t}$.}
   The background color refers to the gap opening mass criterion from \Eq{gap2},  and the grey lines indicate the masses of four Solar System giant planets and $8 \Me$.   
    The color symbols represent the disk setups we have explored in \textbf{Low-viscosity disks},  where the red symbols refer to the circumstances where only Jupiter opens a deep gap (P$0$ is the same as the fiducial run in the main text),   the magenta symbols correspond to the circumstances where both Jupiter and Saturn are in the gap opening regime,  and the orange symbol indicates the circumstance that Jupiter, Saturn, Uranus, and Neptune open gaps while the additional ice giant planet with the lowest mass is in the non-gap opening regime.  
    The values of $\alphat$ and disk aspect ratio parameters can be found in Extended Data \tb{ins}.
    }
    \label{fig:gap}
\end{figure}

\begin{figure}
    \renewcommand\figurename{Extended Data Figure}
    \includegraphics[width=16cm]{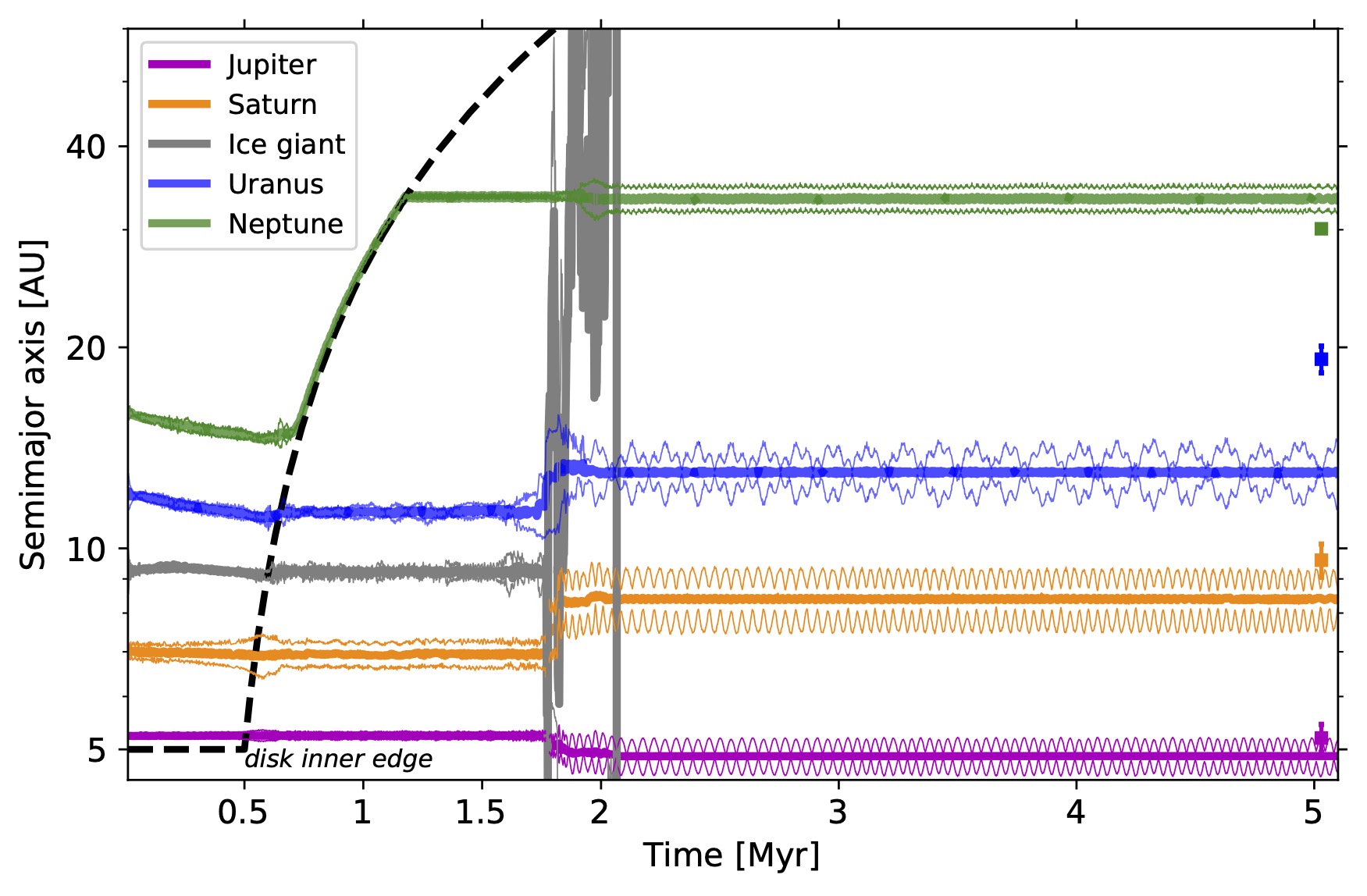} 
    \caption{\textbf{An early dynamical instability triggered by the dispersal of the Sun's protoplanetary disk, assuming that the disk has a low viscosity.}
    The initial system consisted of five giant planets: Jupiter, Saturn, and three ice giants.
    The curves show the orbital evolution of each body including its semimajor axis (thick), perihelion and aphelion (thin).
    The black dashed line tracks the edge of the disk's expanding inner cavity.
    We do not follow the early evolution through the entire gas-rich disk phase, so the onset of disk dispersal is set arbitrarily to be $0.5$ Myr after the start the simulation.
    The semimajor axes and eccentricities of the present-day giant planets are shown at the right, with vertical lines extending from perihelion to aphelion.
    The disk model is adopted from $\textit{run\_B5R\_P4} $ where midplane turbulent strength ($\alpha_{\rm t}{=} 10^{-4}$) is $50$ times lower compared to the example shown in \Fg{illustration}.  The other disk parameters are: $\dot M_{\rm pho} {=}5.5 \times 10^{-10} \Msunyr$, $\taud{=}5.0\times 10^{5}$ yr, and $\vr{=} 42 \auyr$.}
    \label{fig:lowalpha}
\end{figure} 

\begin{figure}
    \renewcommand\figurename{Extended Data Figure}
    \includegraphics[width=16cm]{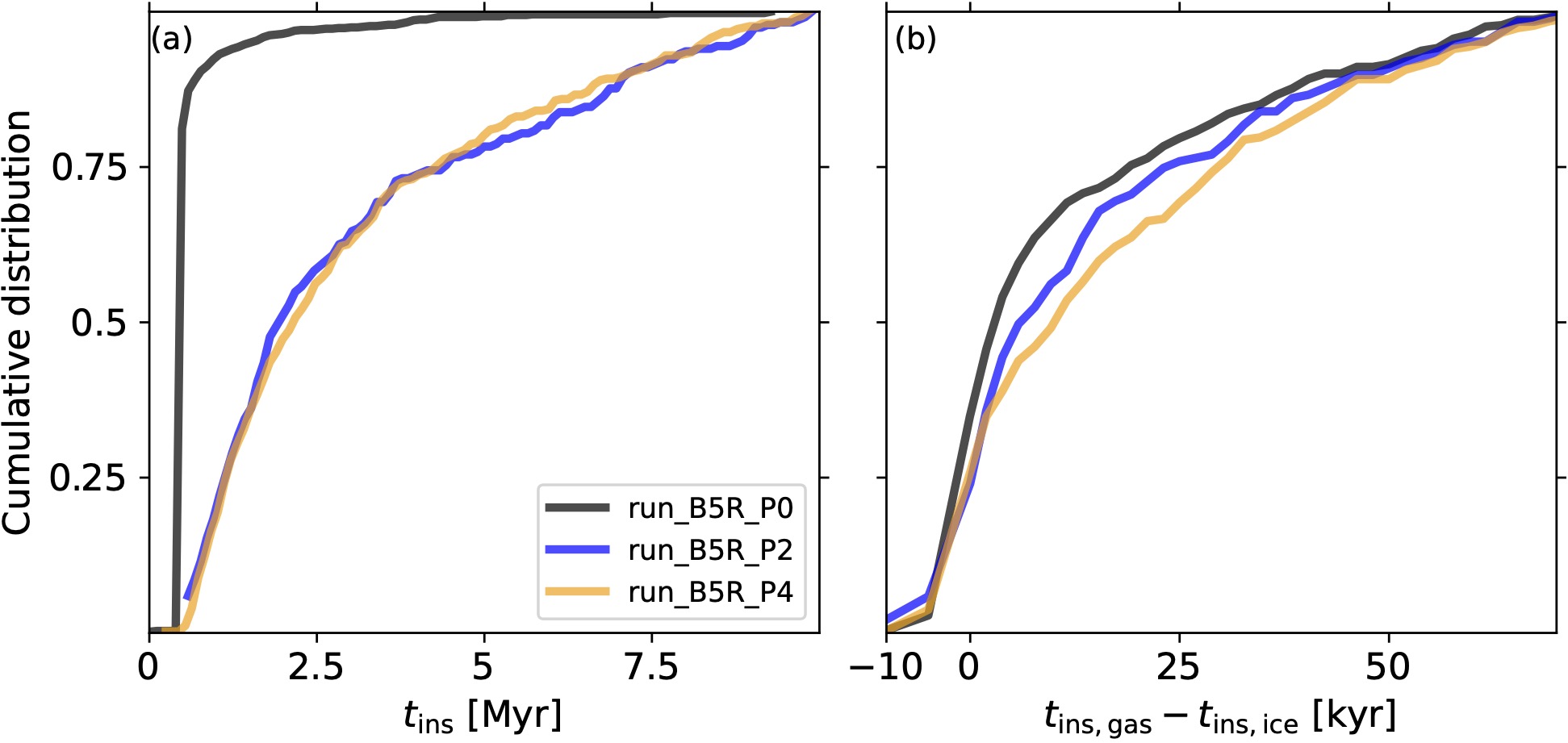} 
    \caption{\textbf{Cumulative distributions of delay times across three different suites of simulations.} On the left, the cumulative distribution of the time to the first instability regardless of which planets are involved. On the right, a cumulative distribution of the time delay between when the ice giant planets undergo orbital instability (typically occurs first), and when the gas giant planets undergo orbital instability.
    The black, blue and orange curves represent the simulations of  $\textit{run\_B5R\_P0} $,  $\textit{run\_B5R\_P2} $ and  $\textit{run\_B5R\_P4} $ in Extended Data \tb{ins}. }
    \label{fig:instability}
\end{figure}

\end{document}